\documentclass[graybox, secnum]{svmult}

\usepackage{mathptmx}       
\usepackage{helvet}         
\usepackage{courier}        
\usepackage{type1cm}        

\usepackage{makeidx}         
\usepackage{graphicx}        
\graphicspath{{./graphics/}}
\DeclareGraphicsExtensions{.pdf,.jpeg,.jpg,.png}
\usepackage[tight,center]{subfigure}
\usepackage{multicol}        
\usepackage{hyperref}        
\hypersetup{colorlinks=false,urlcolor=blue}

\usepackage{upgreek}
\usepackage{gensymb}
\usepackage{siunitx}
\usepackage{hyperref}

\makeindex             

\usepackage[
  backend=biber,
  style=authoryear,
	sorting=nyt,
  uniquename=false,
  uniquelist=false,
  giveninits=false,
  maxcitenames=2,
  maxbibnames=5,
  minbibnames=5,
  block=space,
  giveninits=true,  
  url=false,
  doi=true,
  eprint=true,
  natbib=true
]{biblatex}
\begin{document}

\title*{Proportional counters and microchannel plates}
\author{Sebastian Diebold \thanks{corresponding author}}
\institute{Sebastian Diebold \at Institut für Astronomie und Astrophysik, Eberhard Karls Universit\"at T\"ubingen, Sand 1, 72076 T\"ubingen, \email{diebold@astro.uni-tuebingen.de}}
%
%
\maketitle
\abstract{Developed right at the beginning of the space age in the 1940s, the proportional counter was the first detector used in X-ray astronomy and stayed its workhorse for almost four decades. Although the principle of such a detector seems to be rather simple, over time it underwent considerable performance improvements and the lifetime under orbital conditions was extended tremendously. Particularly the invention of position-sensitive proportional counters provided new and sophisticated methods to discriminate background and thus enabled observations of much weaker sources.\\
A leap forward in position resolution was achieved with the advent of microchannel plate (MCP) detectors in the 1970s. In contrary to gas filled detectors, they provide no considerable energy resolution but feature spatial resolutions reaching down to a few tens of micrometers, fitting ideally the angular resolution of the novel grazing incidence imaging X-ray telescopes upcoming at that time.\\
Even today, both types of detectors are still relevant in space-based astronomy. However, in case of MCPs new developments focus on the far and extreme ultraviolet wavelength range, while the Chandra X-ray observatory is most likely the last mission applying this technology for X-rays. In contrast, compact detectors with gas electron multiplier (GEM) foils and micropattern readout are currently under heavy development for the soft X-ray range, since they allow for the first time to measure polarization in X-rays over a broad energy range.\\
This chapter presents the principles of proportional counters and MCP detectors, highlights the respective performance characteristics, and summarizes their most important applications in X-ray astronomy.}

\section*{Keywords} 
X-ray astronomy, instrumentation, radiation detectors, proportional counters, imaging proportional counters, microchannel plate detectors


\section{\textit{Introduction}}

The history of gas-filled detectors dates back to the onset of nuclear physics at the beginning of the twentieth century with the studies on gas ionization by \citet{Thomson1899} and the first detector implementations by \citet{Rutherford1908}. Shortly after, in 1912, such detectors led to the discovery of cosmic rays when \citet{Hess1912} used them on several balloon flights to investigate the origin of natural radiation.

The proportional counter was introduced in the late 1940s, followed by years of intensive further development mainly driven by its applications in particle physics \citep{Knoll2010}. This phase proceeded until the introduction of NaI scintillation detectors in the 1950s and the first semiconductor detectors in the 1960s. However, the developments on proportional counters continued and lead to the position-sensitive single wire proportional counter (SWPC) and to multi-wire detectors that feature two dimensional imaging capabilities \citep{Fraser2009}.

From the first rocket missions investigating the X-ray radiation from the sun in 1948 \citep{Keller1995}, gas-filled detectors and particularly proportional counters were for almost four decades the workhorse in imaging as in non-imaging soft X-ray astronomy \citep{Pfeffermann2008a,}. In 1962, the first celestial X-ray source outside the solar system was discovered with a proportional counter on a sounding rocket with a remarkable energy resolution of about 20\,\% at 6\,keV. This success hold on and for two decades all main advantages in the field were due to missions applying proportional counters \citep{Bulgarelli2020}. A milestone in the field that has to be mentioned was the first satellite mission fully dedicated to X-ray astronomy: the Uhuru satellite launched in 1970 \citep{Giacconi1971}. It featured a proportional counter sensitive in the band 2--20\,keV with a relatively large sensitive area of $0.084\,\mathrm{m}^2$ and an angular resolution of $1^\circ \times 10^\circ$ FWHM constrained by a collimator. 

According to \citet{Fraser2009}, in X-ray astronomy the main aspects of proportional counters compared to other detectors can be summarized in these three points: 
\begin{enumerate}
	\item relatively large active effective area,
  \item moderate energy and spatial resolution, and
  \item high sensitivity.
\end{enumerate}

Microchannel plates (MCP) -- the second detector technology that is covered in this chapter -- started as a military development for night vision devices, but was declassified and MCPs became commercially available. The main application for the MCP technology is the position-resolved detection of charged particles with a very low energy threshold, but it is also an excellent option for photon detection when an appropriate photocathode for the wavelength range of interest is used. Since the 1980s, MCPs were for many years the main technology for astronomical instruments over the complete ultraviolet (UV) spectral range and they are still competitive in this waveband to silicon-based detectors like CCDs or CMOS. In fact, the best technology for a certain application depends strongly on the individual instrumental requirements and the parameters of the satellite platform. Not only in the UV, but also in the visible and the X-ray bands MCP detectors were successfully applied. However, new instruments and missions with MCP detectors are only proposed in the far (FUV) and extreme UV (EUV, sometimes also XUV).

Considering X-ray astronomy, the key feature of MCP detectors is the unprecedented position resolution. Therefore, it is no wonder that with HRI (High Resolution Imager) \citep{Kellogg1976} flown on the Einstein (HEAO-2) observatory \citep{Giacconi1979} the first modern imaging X-ray telescope employed an MCP detector. The same instrument design was reused after significant further developments for the HRI on ROSAT (ROentgen SATellite) \citep{Pfeffermann1987} and the HRC (High Resolution Camera) of the Chandra X-ray observatory \citep{Weisskopf2002} that is still operable today 23 years after its launch. Furthermore, an MCP detector was also used in the WFC (Wide Field Camera) of ROSAT \citep{Barstow1985}.

Both detector types discussed here work in \textit{counting mode}, meaning that they both register individual photon events. While the proportional counter has an intrinsic (medium) energy resolution, MCP-based detectors usually only resolve energy when used with a dispersive element like a grating and then exploiting their superior position resolution. A second distinguishing feature is the need for a window for the proportional counter to separate the gas volume while an MCP detector can be operated open face. However, the gain in quantum detection efficiency by not having transmission losses in a window for MCP detectors is always over-compensated by the lower intrinsic quantum efficiency compared to a proportional counter.

Meanwhile, proportional counters as well as MCP detectors were almost completely replaced by their principal competitors in X-ray astronomy, namely silicon-based detectors \citep{Knoll2010}. However, there are a few niches for which these technologies are still developed: while MCP detectors are further optimized for the UV and EUV where they can still be competitive to silicon technology depending on the application and the requirements, position sensitive proportional counters are applied recently to measure polarization in X-rays -- a longstanding and scientifically highly interesting topic that is now tackled by several missions.

This chapter presents in Section~\ref{sec:prop_counters} the general concept of proportional counters, their basic parameters, and some considerations on applying them in X-ray space missions. In Section~\ref{sec:imaging_counters} the principle of imaging proportional counters and their application in X-ray astronomy are discussed, including the relevance of micropattern gas detectors for X-ray polarimetry. Section~\ref{sec:MCPs} explains the functionality of MCP detectors and highlights applications in X-ray as well as UV and EUV astronomy. Section~\ref{sec:prospects} concludes with an outlook on the future prospects of both discussed detector technologies.


\section{\textit{Proportional counters}}
\label{sec:prop_counters}

\begin{figure}
	\centering
		\includegraphics[width=\textwidth]{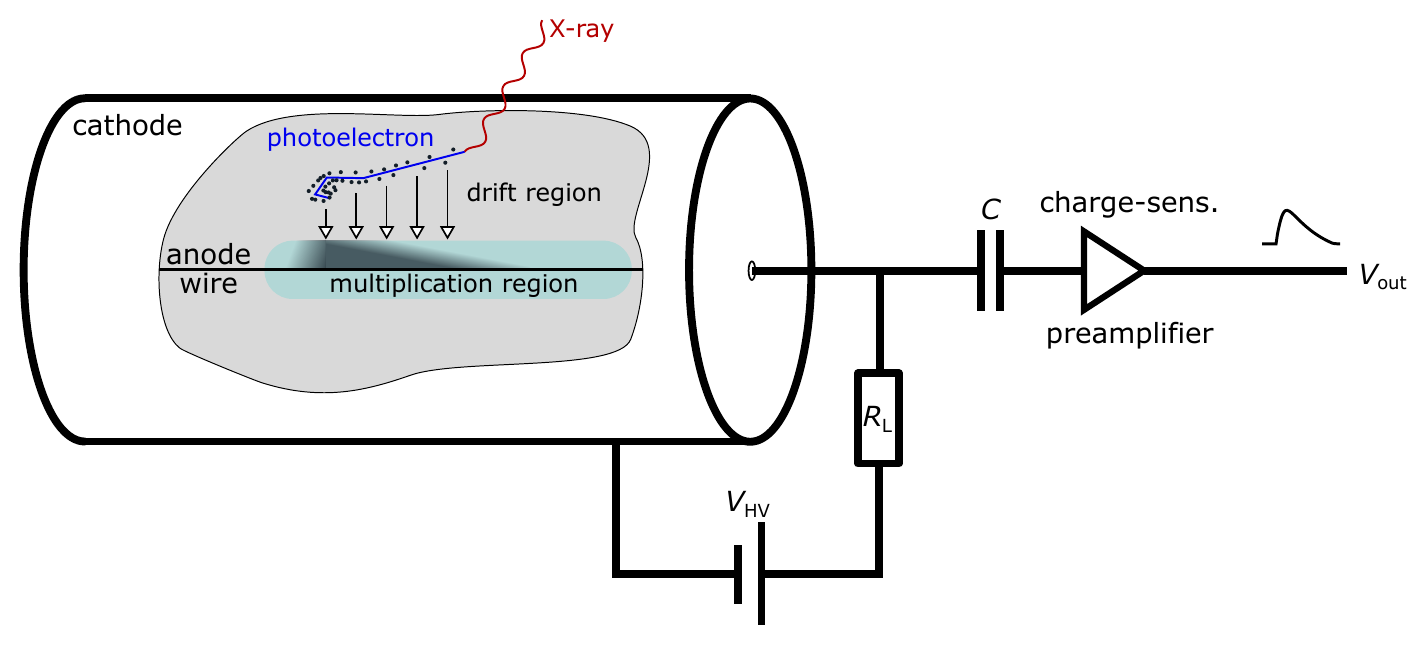}
	\caption{\label{fig:basic_prop_counter}
  Sketch of a proportional counter in the simple radial tube geometry. An X-ray photon interacting with the counting gas generates a photoelectron that loses its energy by ionizing gas atoms. The generated electrons are drifted towards the central anode wire by an externally applied high voltage. When the electrons enter the multiplication region around the anode wire they gain enough energy to ionize neutral gas atoms and initiate a charge avalanche. The signal from the anode wire is read out via a capacitively coupled charge-sensitive preamplifier.}
\end{figure}

The basic principle of a proportional counter is best introduced by looking at the simple radial tube geometry as sketched in Figure~\ref{fig:basic_prop_counter}. A thin wire is mounted coaxially in a cylindrical conductive tube, which is filled with the counting gas and hermetically sealed. While the tube is on ground potential acting as cathode, the insulated wire is connected to a positive high voltage via a load resistance $R_\mathrm{l}$ and forms the anode. When ionizing radiation enters the tube and generates ion pairs in the gas, the high voltage accelerates the electrons towards the anode wire. For the detection of soft X-rays a thin window is needed to allow transmission into the sensitive detection volume.

The field strength is increasing towards the anode wire, subdividing the gas volume in weak and strong electric field regions, usually called \textit{drift region} and \textit{multiplication} or \textit{avalanche region}, respectively. When the electrons gain enough kinetic energy on a mean free path length, collisions with gas atoms can generate secondary electrons in an avalanche process called \textit{gas multiplication} \citep{Knoll2010}. For reading out, the anode wire is capacitively coupled to a charge-sensitive preamplifier that converts the collected charge to a voltage signal. If the applied high voltage and thus the field strength are set properly, the amplitude of the output signal is proportional to the energy deposited by the incident ionizing particle or photon, allowing counting and spectroscopy.

\begin{figure}
	\centering
		\includegraphics[width=\textwidth]{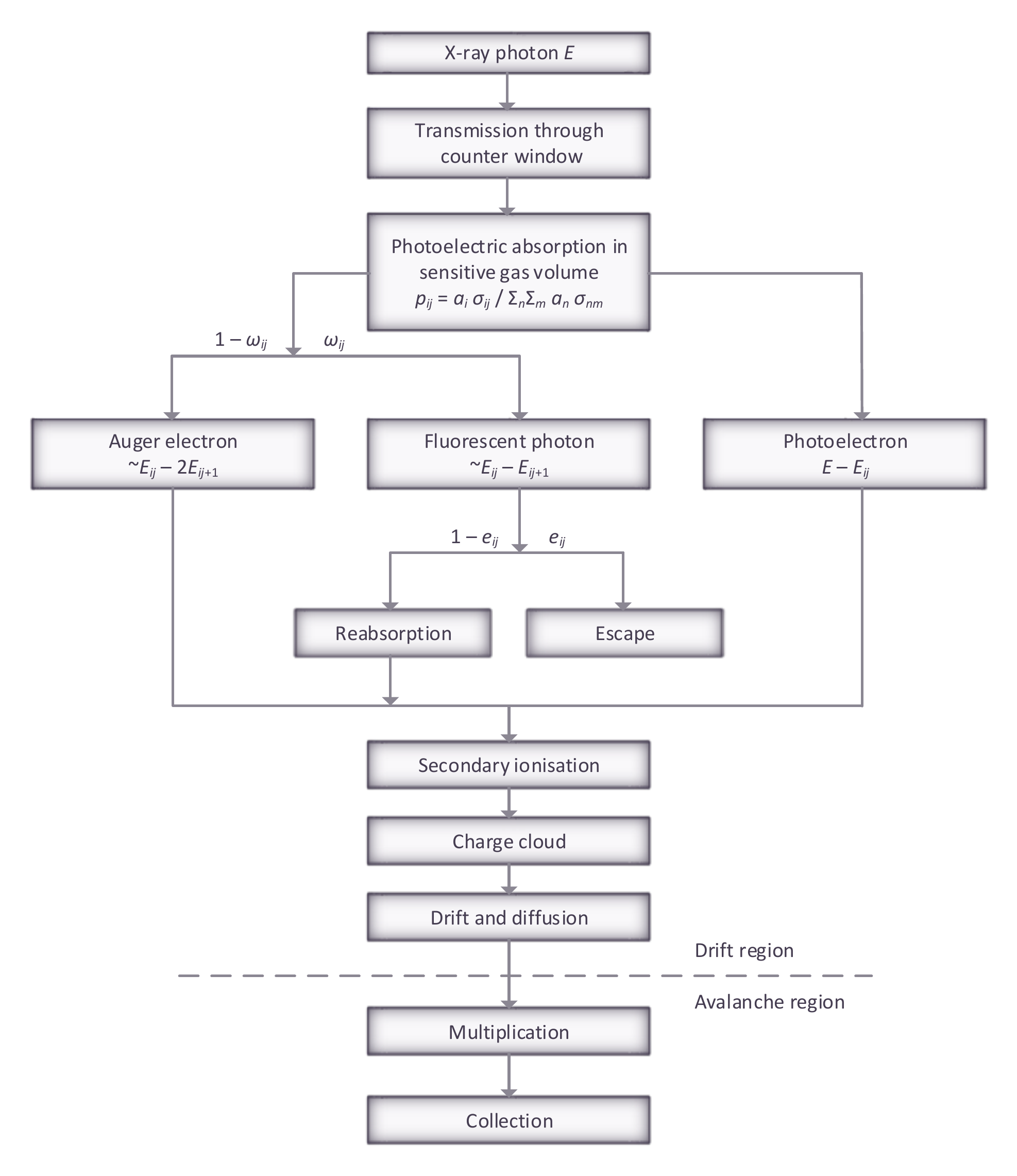}
	\caption{\label{fig:prop_counter_process}
  Schematic overview of the interaction process of soft X-rays in a proportional counter. If the photon energy $E$ exceeds the binding energy $E_{ij}$ of the $j$th shell of the $i$th component of the counting gas mixture, $p_{ij}$ is the probability that a photoelectron is released from this shell. $a_i$ is the fraction of the $i$th atomic component in the mixture and $\sigma_{ij}$ are the respective photoelectric cross-sections. The excited ion can relax either via the emission of an Auger electron or by emitting a fluorescent photon with the probability $\omega_{ij}$. The probability that the fluorescent photon escapes the sensitve counter volume is $e_{ij}$.}
\end{figure}

The complete detection process for soft X-rays in proportional counters is summarized in a schematic overview in Figure~\ref{fig:prop_counter_process}. It should be noted that the photoelectron is not necessarily carrying away most of the energy. A typical example is given in \citet{Fraser2009}: a 6\,keV X-ray photon is absorbed in a mixture of 90\,\% xenon and 10\,\% $\mathrm{CH}_4$. In this configuration 99.83\,\% of the primary interactions are with the xenon L-shell (mean energy 5.0\,keV). These lead to the production of a 1.0\,keV photoelectron and a 4.2\,keV L-fluorescence photon ($14\,\%$ probability) or a $3.4\,\mathrm{keV}$ Auger electron (86\,\%) since the mean M-shell binding energy in xenon is 0.8\,keV.

The individual steps of event detection in proportional counters are discussed in more detail in the following, together with the relevant performance parameters.

\subsection{\textit{Photon interaction via the photoelectric effect}}

The dominating interaction process of soft X-rays below $50\,\mathrm{keV}$ is the \textit{photoelectric effect}. In this process, the photon is absorbed by an electron of an inner atomic shell, usually the K-shell ($n = 1$) if the photon energy is sufficient, and this so-called photoelectron carries away the initial photon energy minus the binding energy. The photoelectric cross section is proportional to $Z^n E^{-8/3}$ with the atomic number $Z$ of the counting gas, the exponent $n$ in the range 4--5, and the photon energy $E$ \citep{Fraser2009}. This implies that for the detection of higher photon energies the mean atomic number $Z$ of the counting gas should be increased to compensate for the enlarged mean free path. Secondly, particularly for the softest part of the spectrum, window materials should be as light weight as possible and built from low-$Z$ material to avoid absorption in the window.

As the photoelectron travels through the gas it loses energy mostly via inelastic collisions with gas atoms forming ion pairs. The typical range of the photoelectron in X-ray counters is 0.1 -- 1\,mm \citep{Fraser2009}. The number $N$ of ion pairs that are generated is linked to the energy $E$ of the photoelectron via the mean energy $w$ for the creation of an ion pair. Without considering any secondary effects the number $N$ of generated electrons by a photoelectron with energy $E$ can be estimated by

\begin{equation}
  N = E/w .
  \label{eq:N}
\end{equation}

$w$ depends on the gas mixture with typical values in the range 25 -- 35\,eV \citep{Knoll2010}. In X-ray counters often noble gases are used, mainly argon with $w = 26.2\,\mathrm{eV}$ for soft X-rays and xenon with $w = 21.5\,\mathrm{eV}$ for harder X-rays \citep{Sipila1976}. Thus, for low-energy X-rays below 10\,keV the number of secondary ion pairs is of the order of just a few hundred.

Since the formation of secondary ion pairs by a photoelectron cannot be described as a series of independent processes, the actual variance $\sigma^2_N$ in the number of generated ion pairs is considerably smaller than expected from Poissonian statistics by a factor $F$ that is called the \textit{Fano factor} \citep{Fano1947}:

\begin{equation}
  \sigma^2_N = FN
  \label{eq:sigma_N}
\end{equation}

Therefore, the variance $\sigma^2_N$ is still proportional to $N$ and the energy resolution $E/\Delta E$ scales with $\sqrt N$. Although an analytical description for $F$ exists, the factor is usually determined experimentally for a specific material \citep{Fraser2009}. Empirical values for $F$ group around 0.15 for typical counting gases \citep{Edgar2011}.

Since the corresponding electrical charge generated for soft X-rays is thus of the same magnitude than the equivalent noise charge at the input of a low-noise preamplifier, an internal amplification via gas multiplication is necessary to perform spectroscopy with a significant \textit{signal-to-noise ratio (SNR)}.

\subsection{\textit{Gas multiplication and energy resolution}}
 
A critical parameter for the operation of a proportional counter is the applied high voltage. Together with the electrode geometry it defines the strength of the electric field and thus distinguishes three operational modes of a gaseous detector:

\begin{enumerate}
	\item \textit{ion chamber mode}, in which the charges scatter only elastically with gas atoms and drift slowly to the electrodes;
  \item \textit{proportional chamber mode}, in which the electrons gain enough energy between collisions to ionize neutral gas atoms and thus form a charge avalanche;
  \item \textit{Geiger counter mode}, in which the acceleration of the electrons is so high that the charge avalanche saturates because the applied field is compensated by the space charge of the generated ions.
\end{enumerate}

\begin{figure}
	\centering
		\includegraphics[width=.8\textwidth]{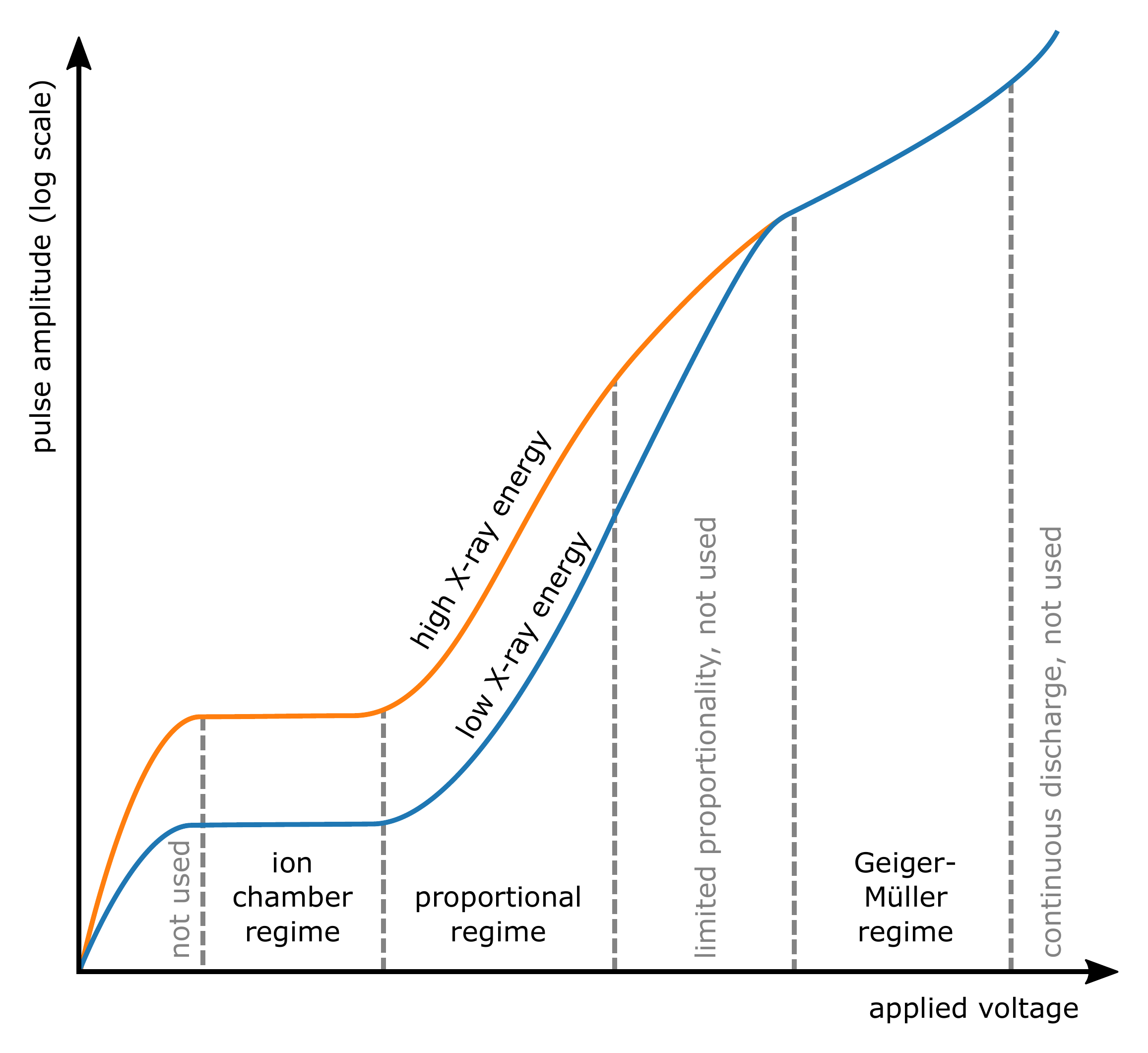}
    \caption{\label{fig:operation_regimes}
    Generic diagram of the different voltage regimes and the according operational modes of a gas counter. In the ion chamber regime the charges drift without multiplication. In the proportional regime at intermediate voltages a linear amplification of the signal by gas multiplication sets in, which becomes non-linear at higher voltage. In the Geiger-Müller regime the amplification completely saturates and the pulse amplitude depends only on the voltage and not on the photon energy any more. At even higher voltages a continuous discharge can be triggered that leads to degradation and ultimately destroys the counter.}
\end{figure}

Figure~\ref{fig:operation_regimes} shows generically the different voltage regimes and how the recorded pulse height depends on the applied high voltage for low and high X-ray energies. The transition region between the proportional and the Geiger-Müller regions is usually not used in scientific detectors.

Only in the proportional chamber mode, the amount of read out charge created by primary X-rays in the few keV range is large enough to allow a detection with discrimination of the energy information. The critical field strength for typical gases at atmospheric pressure to form such a \textit{Townsend avalanche} is of the order of $10^6\,\mathrm{V/m}$ \citep{Knoll2010}. In order to achieve a uniform multiplication that is independent on the exact interaction point in the counting gas, the multiplication region must be much smaller than the total gas volume. Also the geometry, e.g. the wire thickness or the wire distance in a multiwire configuration, is critical to reach a sufficient field uniformity and larger tolerances lead to a decrease of the achievable energy resolution \citep{Fraser2009}.

The derived variance $\sigma^2_e$ of the charge signal and the typical noise contribution from charge amplification $\sigma^2_\text{amp}$ yields a one-sigma energy resolution of

\begin{equation}
  E/\Delta E = w \sqrt{\sigma^2_e + \sigma^2_\text{amp}}.
  \label{eq:res}
\end{equation}

Therefore, with an ideal detector system an energy resolution of 15\,\% can be reached at 6\,keV \citep{Edgar2011}.

At high rates, the energy resolution of a proportional counter can be degraded due to the rather long drift times of the positive ions that are formed close to the anode wire. These ions shield the applied high voltage and thus lead to a local deformation of the electric field, which can affect the gain of subsequent events. It is almost impossible to account for this effect because of its highly localized nature. It could be addressed in position sensitive counters (s. Section~\ref{sec:imaging_counters}), but even there it is usually not corrected.

Critical for the stable operation of a proportional counter is the generation of UV photons by excited gas atoms in the multiplication region as well as during the neutralization of ions at the cathode. UV photons hitting the cathode surface can trigger the emission of additional electrons that can induce subsequent avalanches and eventually a continuous discharge of the counter. This effect is mitigated by the addition of a few percent of a polyatomic \textit{quench gas}, e.g. $\mathrm{CH_4}$ or $\mathrm{CO_2}$, to the usually used noble gases. Such an additive firstly absorbs UV photons and converts the energy into heat, secondly reduces via charge exchange the amount of noble gas ions reaching the cathode, and thirdly increases the drift velocity and thus reduces the influence of gas impurities \citep{Pfeffermann2008a,Sauli1977}.


\subsection{\textit{Detection efficiency and response function}}

For the detection of soft X-rays a thin window of light material is needed to allow transmission into the counting gas down to keV energies and below. Metal windows of beryllium or aluminum allow transmission down to 1.5\,keV and can be used for sealed gas cells. For even lower transmission thresholds, micrometer thick plastics like polypropylene have to be used, but these require a gas supply system to maintain the pressure \citep{Pfeffermann2008a}. The window thickness is usually a compromise between lower energy threshold on the one hand and mechanical stability and gas tightness on the other. Particularly for large windows that enable large sensitive areas, a careful assessment of the applicable forces with respect to the accelerations during launch is necessary. Usually a grid of higher thickness or stronger material is used to support large windows.

The probability for an X-ray photon to pass through the window and interact in the fill gas can be expressed in terms of the optical depths $\tau_\mathrm{w}$ and $\tau_\mathrm{g}$ of window and gas, respectively:

\begin{equation}
  P = e^{-\tau_\mathrm{w}}(1-e^{-\tau_\mathrm{g}})
  \label{eq:interaction_prob}
\end{equation}

The optical depth of a certain material is linearly depending on the mass density $\rho$, the energy dependent mass attenuation coefficient $\mu(E)$, and the material thickness $d$:

\begin{equation}
  \tau = \rho\mu(E)d
  \label{eq:optical_depth}
\end{equation}

The mass attenuation coefficient $\mu(E)$ is simply the interaction cross-section $\sigma(E)$ per mass $m$. As depicted above, X-ray photons primarily interact with inner shell electrons, while the valence electrons are responsible for the chemical bonds. Therefore, the chemistry of the material plays a minor role for the X-ray cross-section and the cross-section for a compound material or gas mixture is simply the sum of the individual cross-sections of the constituents.

\begin{figure}
	\centering
		\includegraphics[width=.8\textwidth]{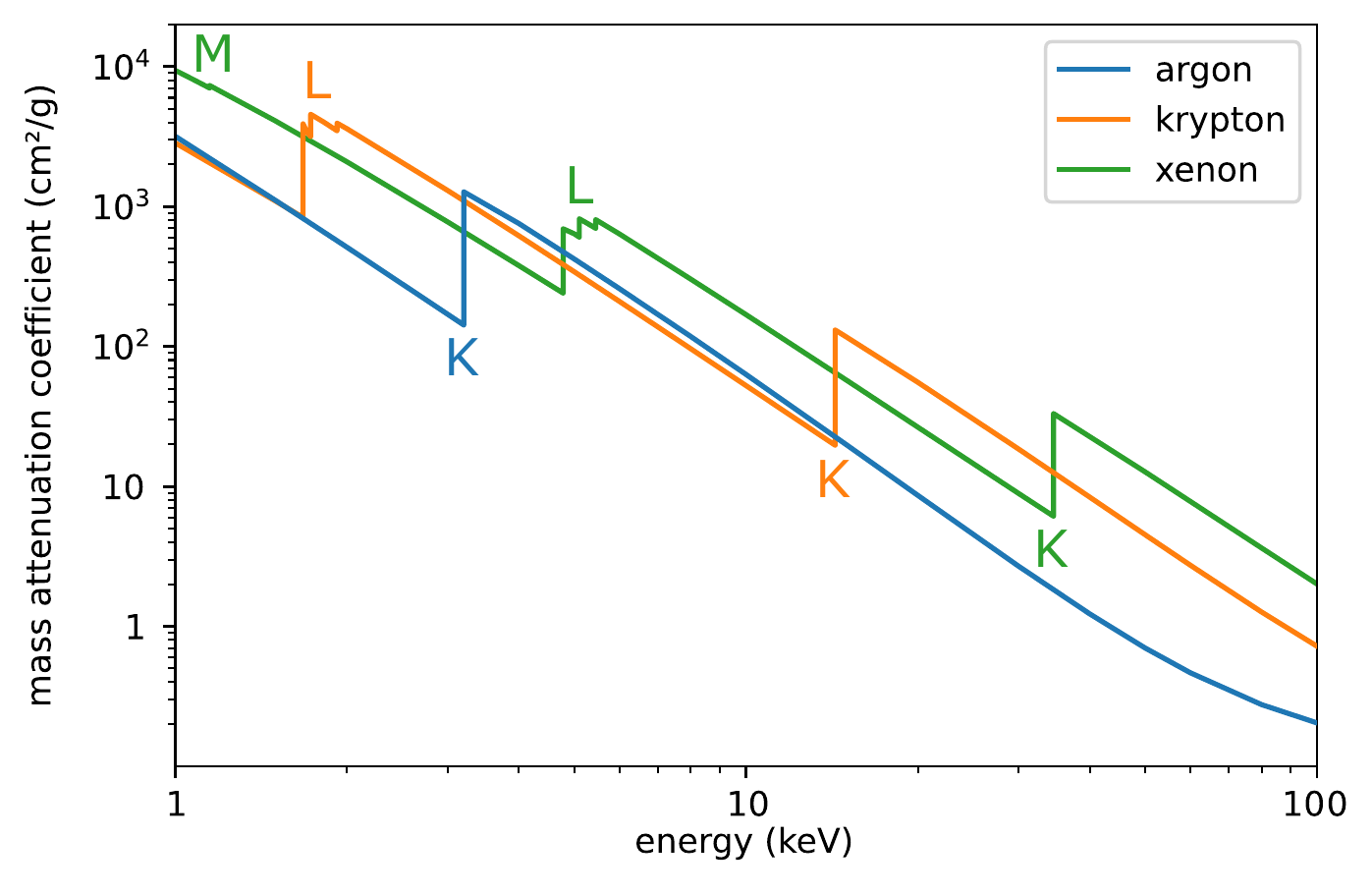}
  \caption{\label{fig:attenuation}
  Mass attenuation coefficients for argon, xenon, and krypton as typical fill gases for proportional counters. A sharp increase of the interaction probability occurs when the photon energy is sufficient to ionize the next inner shell. The respective shell is marked in the plot with M, L, and K. Data obtained from \citet{Hubbell2004}.}
\end{figure}

While $\sigma(E)$ is in general a smooth function, it shows sudden jumps when the photon energy is sufficient to ionize a better bound electron level. Since the $K_\alpha$ energies of common counter gases lie within the energy band of interest ($K_\alpha$ energies of argon, krypton, and xenon are 2.97\,keV, 12.6\,keV, and 29.7\,keV, respectively), the response functions can be complicated \citep{Knoll2010}. While the cross-section and therefore also the mass attenuation coefficient is declining with rising X-ray energy, a steep increase is seen when reaching the K-shell energy (see Figure~\ref{fig:attenuation}).

Furthermore, the characteristic X-ray radiation \textit{(fluorescence)} generated by the photoelectric absorption carries away a significant portion of the primary photon energy. Events in which these X-rays are not absorbed in the gas lead to an additional \textit{escape peak} in the pulse height spectrum, which is shifted from the \textit{full-energy peak} to lower energies by the $K_\alpha$ energy. Additional peaks in the spectrum might originate from the characteristic lines generated when X-rays are absorbed in the entrance window or in the walls. Besides their higher X-ray transparency, low Z window materials and surface covers are thus also favorable to minimize this contribution \citep{Knoll2010}.

Due to all these effects, the conversion from photon energy to mean pulse height -- often referred to as \textit{gain} -- can be non-linear with jumps and slope changes, particularly near absorption edges. Even a monochromatic source produces a complicated pulse height spectrum with several features. Thus, a direct estimate of the true photon spectrum from measurement data is not possible and usually a \textit{response matrix file (RMF)} is used to analyze the data of scientific observations. Such an RMF contains the measured probabilities for the different possible pulse heights over the complete photon energy range in which a detector is sensitive \citep{Edgar2011}.

\subsection{\textit{Time resolution, dead time, and rate limitation}}

The time resolution of a proportional counter is dominated by the drift time between the interaction point of the photon and the avalanche region and thus depends particularly on the size of the drift region. Since drift velocities of electrons in gases are high and reach $10^6$ -- $10^7\,\mathrm{cm/s}$ also for moderate electric fields, ultimate time resolutions below $1\,\mathrm{\mu s}$ are reachable \citep{Pfeffermann2008a}.

However, also in a fast detector system an event can only be recorded correctly when a sufficient time interval has passed after the previous event during which the detector system has attained its \textit{ground state}. All produced charges need to be collected -- for a proportional counter this is dominated by the drift time of the positive ions to the cathode -- so that the electric field is completely recovered and the output voltage has settled at the baseline value. Further, the electronic processing of the previous event needs to have finished setting the analog-to-digital converter (ADC) in a ready state for the next event.

In case of a second photon event before full recovery we speak of \textit{pile-up} or \textit{dead time}. The probability for such events increases with increasing count rate. Pile-up means that two or more photon events are so close together in time that the electronics cannot distinguish between them. The signals of these events add up and only a single event with about the sum of the energies is recorded. Due to the distorted waveform, e.g. a longer rise time or a multiple peak structure, some detector systems can filter at least a fraction of pile-up events \citep{Usman2018}.

If the detector is intrinsically fast like a proportional counter, it is possible to use a fast trigger technique to detect incoming events and to close the electronics for a defined time to not process new events until the detector output has certainly reached its baseline value. In case a second event arrives during dead time, the dead time window is restarted. This so-called \textit{paralyzing dead time} mode leads to the situation that at very high count rates the detector hardly processes any events and thus sets a hard limit on the maximum count rate. However, to not excessively degrade a proportional counter the actual rate during observations should be restricted to lower values, typically well below $1000\,\text{/s}$.

In order to know the real exposure time -- e.g. for the correction of the count rate or the calculation of the photon flux from a particular source -- the dead time needs to be subtracted from the exposure time. The resulting time is then called the \textit{live time}, in which the detector was able to register events. Due to the different mechanisms creating dead time, the exact determination can be quite complex and usually only a partial correction is possible \citep{Usman2018}.

However, various techniques are applied to determine the dead time: a periodic pulser signal can be injected in an unused part of the spectrum and the fraction of recorded to sent pulser events yields the fraction of live time to exposure time. Another method uses the typical event length and cuts extended events that are likely pile-up. From the number and time structure of such events also the live time can be deduced. 

It is important to note here that also background events induce dead time in a detector system, even if they can be identified and rejected, thus reduce the live time of observations and the instrumental throughput. Therefore, a thorough simulation of the instrument in the orbital radiation environment is crucial to assess the performance and whether the scientific goals of a mission can be reached (cf. Section~\ref{sec:op_space}).

\subsection{\textit{Operation in space: background and lifetime}}
\label{sec:op_space}

An X-ray detector in space must withstand the highly variable and partially quite harsh orbital radiation environment. On the one hand, the exposure to charged particle and excessive photon fluxes degrades the detector performance -- particularly the energy resolution but also the sensitivity might be affected -- and, on the other hand, external and internal radiation sources create background in scientific observations. This list covers the most important sources of radiation in orbit besides the astrophysical target:

\begin{itemize}
	\item \textit{cosmic rays} are charged particles with a wide energy spectrum up to the highest energies;
  \item \textit{soft protons} are mostly of solar origin, can reach X-ray detectors thorough optics and filters, are highly variable, and often deposit energy in the instrumental X-ray band \citep{Diebold2015};
  \item \textit{trapped charged particles} in the radiation belts of the Earth;
  \item \textit{fluorescent X-rays} emitted from surrounding materials of the detector;
  \item \textit{radioactive impurities} in the detector, e.g. not completely radio-pure counting gas or trace elements in the housing and sealings; and
  \item the \textit{diffuse X-ray background}.
\end{itemize}

Depending on the orbit of the mission the influence of these individual components on the instrument varies strongly. It is interesting to note that for most astrophysical X-ray sources and the typical instruments the background rate exceeds the X-ray event rate \citep{Pfeffermann2008a}. Therefore, methods for background detection and rejection via event selection are applied in all modern X-ray instruments.

In case of \textit{minimal ionizing particles (MIPs)} that deposit about $1.5\,\text{MeV/cm}$ the size of a proportional counter is usually large enough so that the deposited energy is above the upper threshold of the detector and thus the event is rejected. Another internal but more sophisticated method to suppress background events exploits the difference in the geometric shape of the charge cloud induced by a particle and by a photon. While photon events produce a more localized, point-like ionization cloud, charged particles leave an extended ionized track. This leads to different rise times of the output signals and thus provides a handle to discriminate particle events even when the energy is within the band of interest.

A very efficient method to identify background events is the use of an \textit{anticoincidence detector} -- sometimes also called \textit{veto detector} -- that ideally completely surrounds the actual detector volume and leaves only the entrance window for the X-rays uncovered. In case of simultaneous events in both the actual and the anticoincidence detector, the events are either directly rejected or marked as possible background, because of the high probability being triggered by a non-X-ray interaction. If the anticoincidence detector shares the same gas volume with the main detector, also fluorescent X-rays generated by charged particles in the walls can be absorbed with high efficiency in the anticoincidence region and thus do not participate to the overall background. However, this works only effectively if the gas column density of the anticoincidence region is sufficiently high.

By combining these methods overall background rejection efficiencies of up to 99.6\,\% have been realized even for large area proportional counters for X-ray astronomy \citep{Fraser2009}. In the imaging proportional counter PSPC (Position Sensitive Proportional Counter) of ROSAT a rejection of even 99.85\,\% has been reached \citep{Pfeffermann2003}.

But charged particles not only produce background in the detector, they can even lead to degradation or failure. Two major mechanisms affect proportional counters in space: the interaction of heavy charged particles and the aging of the quench gas. Heavy ionizing particles hitting the detector can deposit up to four orders of magnitude more energy in the sensitive gas volume than X-rays. Therefore, the counter design and the operation parameters must be carefully selected that no permanent discharge or spark discharge can be triggered by such events, which can lead to permanent failure of the detector, e.g. by destroying the anode wire.

A permanent aging of proportional counters is initiated by the cracking of quench gas during the normal operation. In this process hydrocarbons like $\mathrm{CH_4}$ lead to the deposition of polymerization products on the internal electrodes that induce a progressing gain degradation by effectively increasing the radius of the anode wire and eventually a permanent discharge, the so-called Malter effect. Gas composition and purity, the strength of the electric field, as well as the materials used for wires, housing, and sealing affect the rate of this degradation \citep{Pfeffermann2003}. $\mathrm{CO}_2$ has been proposed as an alternative quench gas that completely avoids polymerization and produces only less harmful carbon depositions \citep{Ramsey1994}.


\section{\textit{Imaging proportional counters}}
\label{sec:imaging_counters}

Even for the simple tube geometry of a proportional counter, position sensing of the event in the direction of the anode wire is possible. Since the drift field is radially, the position of the avalanche is a good indication where the primary photon interaction happened. The most common technique uses charge division, for which the anode wire needs a significant resistance and the charge is read out on both ends. By dividing the amount of charge recorded at one end by the total charge the position information is determined. The total charge holds still the information of the primary energy.

An alternative position sensing technique for the linear counter geometry exploits the rise time of the pulse which is longer when the charge is collected on a resistive anode wire far from the preamplifier, since the wire can be regarded as an $RC$ transmission line because of its finite capacitance to the cathode. By reading out at both ends of the wire, the achievable position resolution can be improved considerably by comparing the respective rise times. Although theoretical considerations show that the charge division technique should yield position resolutions by up to a factor of two higher, in practical applications the rise time sensing turned out to be superior due to its simpler realization in analog electronics \citep{Fraser1981,Fraser1981a}. Therefore, this was also the technique of choice for the application in detectors for X-ray astronomy, first in one dimension and later also in two dimensions \citep{Fraser2009}.

While position sensing in linear counter geometries finds a meaningful application e.g. in the readout of a dispersive spectrometer, two dimensional designs are needed for imaging applications. A simple form is to use several anode wires each with a dedicated readout channel, a so-called multiwire proportional counter. For a two dimensional position determination of the photon event, either the wires can be read out on both sides following one of the schemes presented above, or a grid of crossed wires can be used. However, the position resolution in the two axis will be significantly different with different systematics.

\begin{figure}
	\centering
		\includegraphics[width=.8\textwidth]{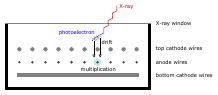}
  \caption{\label{fig:imaging_principle}
  Concept of an imaging proportional counter. The charges produced by an X-ray event are drifted through a cathode wire grid to the multiplication region. The charge cloud is read out at one of the anode wires, producing a negative signal that is used to determine photon energy and arrival time. The position information is gained from the positive signals induced in the two layers of cathode wires, which are oriented perpendicular to each other.}
\end{figure}

The concept of an imaging proportional counter that reaches the same position resolution in both directions is shown in Figure~\ref{fig:imaging_principle}. A layer of thin anode wires is embedded between two layers of cathode wires that are oriented perpendicular to each other. The negative charge signal read out from one of the anodes is used to reconstruct energy and timing information of the X-ray event while the position is determined from the positive signal induced in the cathode layers. While in principle the same energy resolution as for a single wire counter can be reached, the tolerance of the anode wire pitch must be of the order of micrometer to reach a uniform gain over the whole detector \citep{Pfeffermann2008b}.

\subsection{\textit{Position resolution}}

The achievable position resolution in an imaging proportional counter depends on several components that are all independent from each other and hence add up to the spacial accuracy $\sigma_\text{x}$ of the detector system \citep{Pfeffermann2008b}:

\begin{itemize}
	\item $\sigma_\text{a}$: Focused X-rays from the mirror system hit the detector not normal to the surface and the actual absorption depth varies from photon to photon and with energy. This can be partially mitigated by displacing the focus plane inside the gas volume.
  \item $\sigma_\text{r}$: The created charge cloud has a finite size depending on the ranges of the photoelectron and the fluorescent photon or Auger electron.
  \item $\sigma_\text{d}$: During the drift the electron cloud widens via lateral diffusion.
  \item $\sigma_\text{e}$: The detection accuracy of the position of the electron cloud depends on its geometrical size before multiplication and the number of primary electrons. Particularly for low photon energies electronics noise degrades the resolution further.
\end{itemize}

\begin{equation}
  \sigma_\text{x}^2 = \sigma_\text{a}^2 + \sigma_\text{r}^2 + \sigma_\text{d}^2 + \sigma_\text{e}^2
\label{eq:pos_res}
\end{equation}

In case of soft X-rays, electron statistics dominate the overall position resolution. Therefore, the position resolution scales as the energy resolution with $1/\sqrt{E}$.

\subsection{\textit{Imaging proportional counters in X-ray astronomy}}

Detectors based on this geometry have been flown on several X-ray missions. In the IPC (Imaging Proportional Counter) of the Einstein (HEAO-2) mission \citep{Giacconi1979} the cathode wires in each layer were connected in a meander-like pattern to form two $RC$ delay lines. The event position was reconstructed by comparing the signal shape at both ends of the delay lines. The longer the signal traveled before the readout the longer the rise time became. This concept reached in the IPC instrument a position resolution of about $1\,\mathrm{mm}$ FWHM \citep{Gorenstein1981}.

For the PSPC (Position Sensitive Proportional Counter) instrument of ROSAT the readout scheme was refined by using individual channels for strips of four adjacent cathode wires. Since the charge signal of an X-ray event was distributed on several strips, a center of gravity calculation allowed a position resolution smaller than the strip width. Eventually, the PSPC reached a spatial resolution of $230\,\mathrm{\mu m}$ \citep{Pfeffermann2003}. Additionally to the background suppression with a system of five anti-coincidence detectors, the charge distribution over the strips was also exploited to reject particle-induced events.

Although not an imaging detector system, the PCA (Proportional Counter Array) of the RXTE (Rossi X-ray Timing Explorer) mission \citep{Bradt1991} used a similar geometrical concept of individual anode wires embedded in cells surrounded by grounded wires. While photon events produce signals only in one of the cells, an energetic charged particle would traverse several cells and can thus be discriminated. Additionally, to improve background rejection, a veto layer around the sensitive detector volume is formed within the same gas volume. In fact, these possibilities for background detection and rejection is one of the attractive features in applying proportional counters as well as imaging proportional counters in X-ray astronomy, particularly for the realization of large sensitive areas.

\subsection{\textit{Micropattern gas detectors and X-ray polarimetry}}
\label{ssec:micropattern-gas}

With micropattern gas detectors a complete new kind of gaseous detectors came up at the beginning of the 1990s. Due to considerable advances in manufacturing techniques like photolithography, selective etching, and laser machining, the wire structures necessary for the strong field multiplication region of a proportional counter could be replaced by microscopic structures on glass or plastics substrates \citep{Knoll2010,Pinto2010}.

JEM-X on Integral \citep{Lund2003} was the first instrument dedicated to X-ray astronomy that applied this concept in its position-sensitive \textit{microstrip gas chamber}, a configuration proposed first by \citet{Oed1988}. Fine anode structures interleaved with wider cathodes are produced on the front surface of a partially insulating glass substrate. The backside of the substrate features a second layer of cathodes that are perpendicular to the front ones. Therefore, geometry and readout are similar to a classical imaging proportional counter as described above, but in a miniaturized, highly ruggedized form. One of the main advantages is a superior gain uniformity due to low tolerances and a high stability of the anode geometry.

A recent and novel application of micropattern \textit{gas pixel detectors (GPD)} is polarimetry in the X-ray domain. The measurement of polarization is a long-standing topic in X-ray astronomy, but different to other wavebands as radio or visible are polarimetric instruments in the X-ray domain complex and difficult to realize. Detecting the degree and direction of polarization yields the possibility to constrain the geometry of the emission region without spatially resolving the source. A first approach with a considerable sensitive area on a satellite was already made with OSO-8 in 1975 by using Bragg reflection off of crystals for which the cross-section is dependent on the polarization of the incoming radiation. However, this principle works only in a narrow, almost monochromatic band, and also due to the small effective area of the instrument the only detected source with a finite polarization was the crab pulsar \citep{Weisskopf1976}.

After this first try, the topic was not touched for decades until GPDs enabled a quite sophisticated principle to measure polarization: The emission direction of a photoelectron generated via the photoelectric effect is correlated with the direction $\phi$ of the electric field of the absorbed photon. Therefore, for a polarized source the number of detected events varies with $\cos^2\phi$. \textit{Gas electron multiplier (GEM) foils} \citep{Sauli1997,Sauli2016} in combination with micropattern anodes directly attached on a readout ASIC (application specific integrated circuit) allow to record the ionization track of the photoelectron. The gas cell with about 1\,mm of dimethyl ether and helium is sealed with a beryllium window transparent to X-rays. More details on this detector concept can be found in the Chapter \textit{Gas pixel detectors for polarimetry} within this handbook.

The first mission to apply a gas pixel detector in a polarimetric instrument is IXPE (International X-ray Polarization Explorer), launched in 2021 and in operation at the time of writing \citep{Weisskopf2022}. With eXTP (enhanced X-ray Timing and Polarimetry) the next mission with a larger grasp that applies the same technology is already being prepared \citep{Zhang2018}. Both missions are presented in detail within this handbook in the section \textit{X-ray Missions}.


\section{\textit{Microchannel plate detectors}}
\label{sec:MCPs}


\subsection{\textit{Channel electron multipliers}}
\label{ssec:CEM}

\begin{figure}
	\centering
		\includegraphics[width=.7\textwidth]{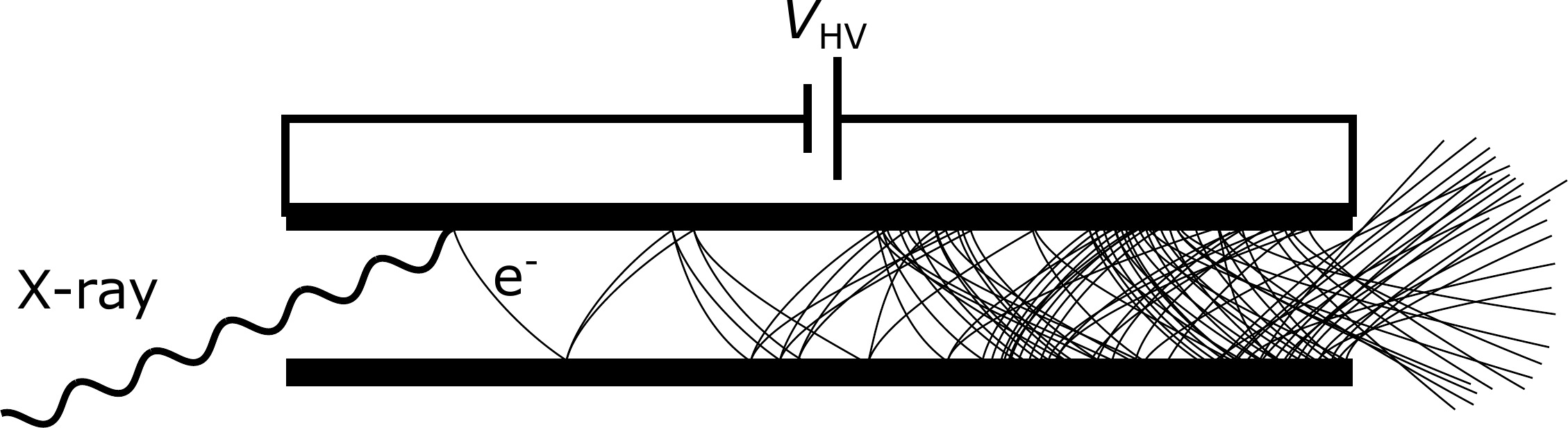}
	\caption{\label{fig:cem}
  Cross-section of a CEM. An incoming X-ray or charged particle releases an electron when impacting the channel wall. The electron is accelerated by the applied high voltage and generates an avalanche of additional electrons. The exiting charge cloud consisting of up to more than $10^8$ electrons can be detected on an anode.}
\end{figure}

In order to explain the principle of microchannel plates we first take a look at a simpler detector that applies the same principle of charge multiplication: the (straight) \textit{channel electron multiplier (CEM)}. These devices are somewhat similar to conventional photomultiplier tubes, but usually smaller in size and with a continuous dynode structure instead of discrete gain stages. The principle is shown in Figure~\ref{fig:cem}. Incoming ionizing radiation that hits the inner wall of a tube releases one or more electrons. In case the primary was a photon we speak of photoelectrons; for particles the secondaries are called delta electrons. A high voltage applied along the tube accelerates the electrons that generate additional secondaries when striking the wall again. In this avalanche process a charge cloud forms that can be detected on an anode. The \textit{gain} -- typically defined as the mean number of electrons that are in total released in one event -- is in the range of $10^4$ up to more than $10^8$. A semiconducting coating on the tube walls recharges after each event with a time constant depending on the resistivity. Since the mean free path under atmospheric conditions would not be sufficient to trigger secondary electron emission, these devices can only be operated under vacuum.

Typical tube diameters of channel electron multipliers are of the order of millimeters up to several millimeters. First imaging detectors that apply this concept of charge multiplication were produced by the formation of two-dimensional arrays of CEMs. However, this intricate and complex procedure led only to a course grid with a limited sensitive area. A leap in spatial resolution of such arrays was achieved when producers were able to form millions of channels with a diameter on the scale of tens of micrometer: \textit{the microchannel plate}.

\subsection{\textit{Microchannel plates}}

Microchannel plates (MCPs) started as military technology for night vision devices, but were later declassified and applied in many particle and photon detectors \citep{Lampton1981}. The principle of multiplying charge is the same as in the CEM (s. Section~\ref{ssec:CEM}), but an MCP consists of millions to billions of microscopic multiplier channels arranged in a regular pattern \citep{Wiza1979}. The scales of the channel diameter typically range from a few micrometers to a few tens of micrometers with a pitch only slightly larger than the diameter. This channel micro-structure sets the ultimate limit on the achievable spatial resolution of an MCP detector, which is rather easily reached with decent readouts. Today, MCPs are available in a wide variety of formats and sizes: plate diameters range from the centimeter scale up to a few ten centimeter, and while the most common shape is still round, also squared and hexagonal formats as well as plates with central holes -- e.g. for a primary beam -- are available from a handful of companies. Of particular interest for astronomical instruments are curved plates that are more difficult in production and accordingly costly but can match better the shape of the focal surface of an instrument -- most prominent example is a curved linear detector for a Rowland spectrograph. It should be noted here that although the market share of astronomical applications is minute compared to other fields, many developments in the MCP technology were initiated by the demands of astronomers \citep{Fraser2009}.

For the fabrication of conventional MCPs, a macroscopic tube of lead glass is mechanically supported by inserting a rod of soluble glass. The assembly is drawn through an oven after which the thickness is reduced to about 1\,mm. The fibers are cut and assembled in hexagonal stacks, which are again heated and drawn, cut, and finally assembled in a hexagonal capsule. The tubes that have now reached their desired diameter of a few micrometer are fused together under vacuum. The capsule is then sliced, polished, and the edges are grinded to the desired final shape, usually to a circle. The inner support is removed by etching and the channels are treated under a hydrogen atmosphere to produce a semiconducting surface layer until the desired resistivity and secondary electron emission yield are reached. Finally, metal electrodes are deposited on both faces of the plates for electrically contacting and applying a high voltage across the channels.

A rather new development are borosilicate MCPs that are manufactured by the hollow-tube technique without the need for a soluble support inside the channels \citep{Ertley2018}. These MCPs are functionalized with atomic layer deposition (ALD) by growing first a resistive layer and on top a layer with high secondary emission yield, i.e. a low electron work function \citep{Gebhard2019}. These layers formed with ALD have a high uniformity throughout the channel and their composition is more stable over time than the surface layer of conventional plates. While for lead glass MCPs a considerable degradation of the gain and changes in the resistivity appear typically after an extraction of $0.1\,\mathrm{C/cm^2}$ due to the electron bombardment of the surface layer of the channels, ALD MCPs have proven to be stable up to an extracted charge of at least $7\,\mathrm{C/cm^2}$ \citep{Popecki2016}. Furthermore, since lead glass contains traces of radioactive isotopes -- particularly the beta-instable $^{40}\mathrm{K}$ -- conventional plates have a typical dark count rate of $>0.25\,\mathrm{/cm^2/s}$, while borosilicate MCPs with their low level of intrinsic radioactivity show a considerably reduced dark count rate of $<0.05\,\mathrm{/cm^2/s}$ \citep{Ertley2018}.

\begin{figure}[t!]
	\centering
	\subfigure[Conventional MCP, 50\,mm diameter]{
  	\includegraphics[width=.47\textwidth]{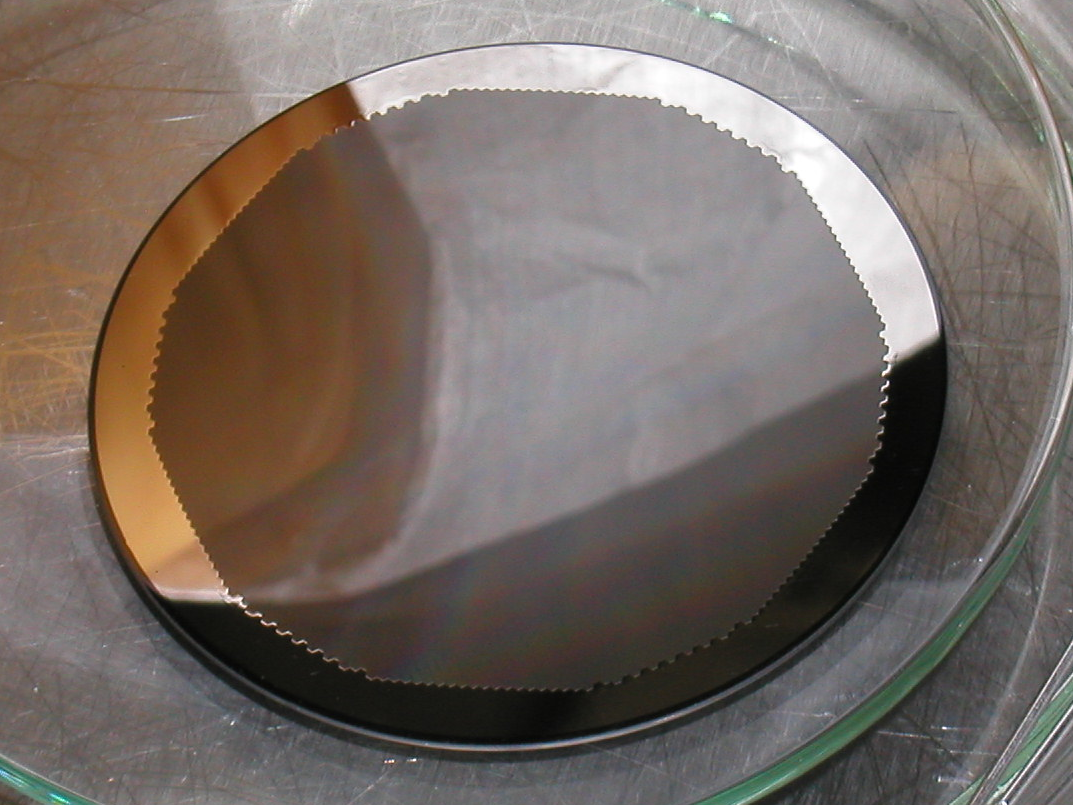}
		\label{fig:MCP_conv}
	}
  \subfigure[ALD MCP, 50\,mm diameter]{
  	\includegraphics[width=.47\textwidth]{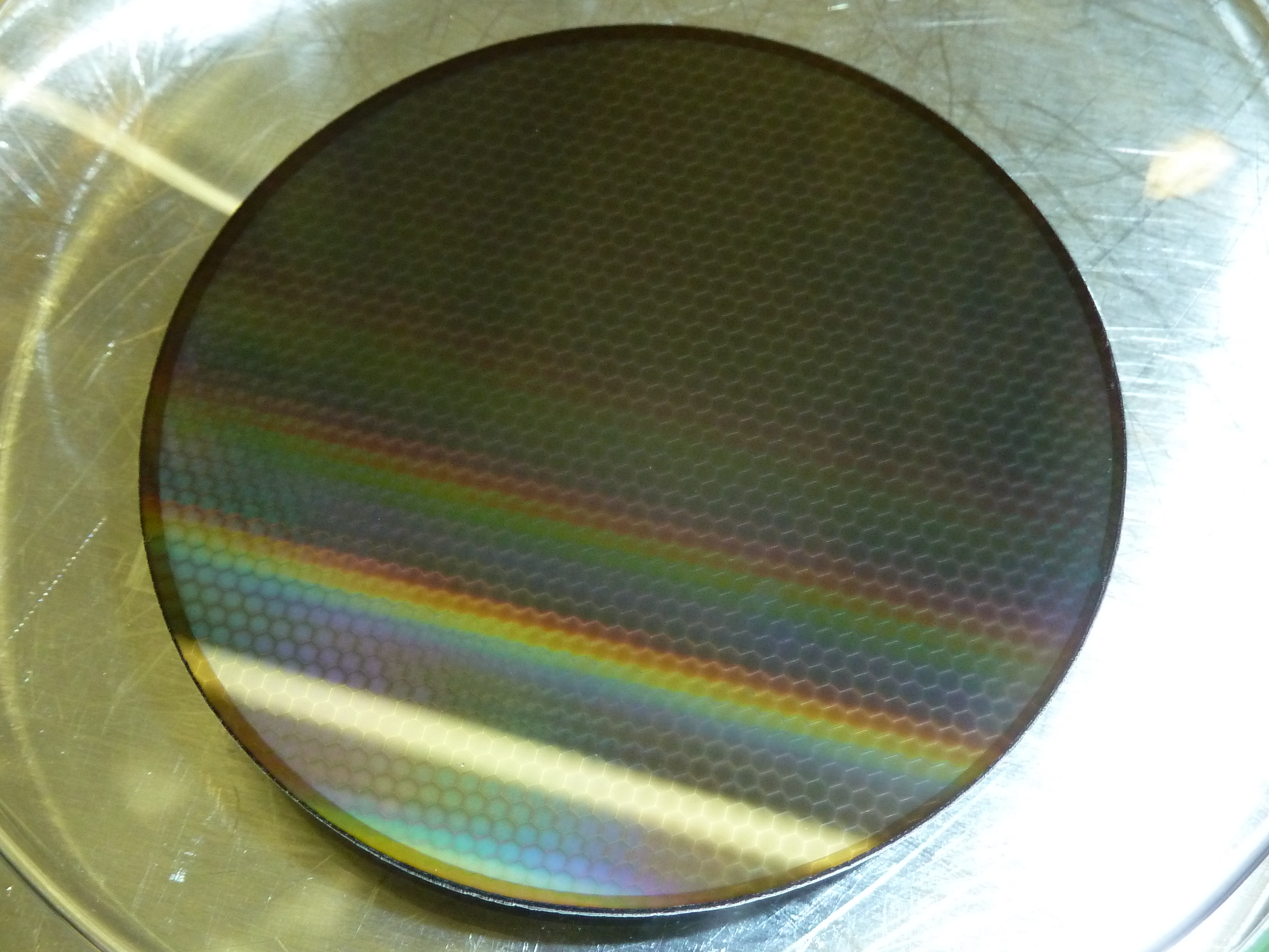}
		\label{fig:MCP_ALD}
	}
	\subfigure[ALD MCP, $20 \times 20\,\mathrm{cm^2}$ square]{
  	\includegraphics[width=.6\textwidth]{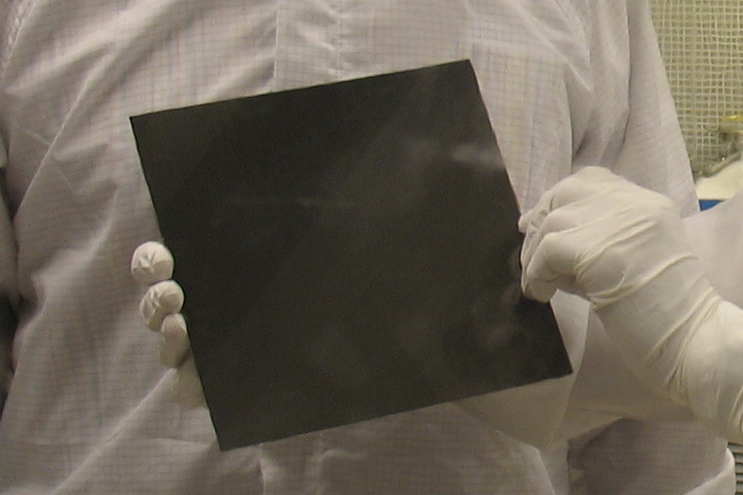}
		\label{fig:MCP_LAPPD}
	}
	\caption{\label{fig:MCPs}
  Photographs of various commercially available MCPs: (a) Conventional round lead glass MCP (Photonis, 50\,mm diameter, 12° bias angle, 0.6\,mm thickness, $10\,\upmu\mathrm{m}$ channel diameter, $l/d$ 60:1, $12\,\mathrm{\upmu m}$ pitch) (b) Round borosilicate ALD MCP (Incom, 50\,mm diameter, 13° bias angle, 0.6\,mm thickness, $10\,\upmu\mathrm{m}$ channel diameter, $l/d$ 60:1, $12\,\mathrm{\upmu m}$ pitch) (c) Large square borosilicate ALD MCP (Incom, prototype for Gen I LAPPD, $203 \times 203\,\mathrm{mm^2}$, 0.6\,mm thickness, $20\,\upmu\mathrm{m}$ channel diameter, $l/d$ 30:1)}
\end{figure}

Photographs of a conventional lead glass MCP, a borosilicate ALD MCP of typical size, as well as an exceptionally large format borosilicate ALD MCP are shown in Figure~\ref{fig:MCPs}.

\subsection{\textit{Operation of MCPs in detectors}}

A key parameter of an MCP is the gain $g$, i.e. the number of electrons in the emitted electron cloud. The fact that the gain of an MCP channel -- keeping voltage and channel surface conditions fixed -- is mainly determined by the dimensionless length-to-diameter ratio $l/d$ allowed for the miniaturization of CEMs to MCPs. Typical $l/d$ ratios of MCPs are in the range 40/1 -- 60/1 because such plates show a minimum spatial variation of the gain. However, for photon-counting applications a larger $l/d$ is usually an advantage and thus also thicker plates up to 175/1 are available \citep{Fraser2009}.

The resistance over an MCP is the result of the individual channel resistances connected in parallel. Typical values are $>100\,\mathrm{M\upOmega}$ with the actual value following roughly a scaling with $l/A$, where $A$ is the plate area. The applied voltage generates a finite current through the semiconducting channel walls and thus heats the material. Since the resistance has a non-ohmic behavior and decreases with increased bias voltage, a thermal runaway can be triggered when the critical current is exceeded, leading to the destruction of the plate. From this point of view a larger resistance seems favorable. However, on the other hand, the resistance influences the count rate capability since at high count rates the positively charged channel walls might not fully regenerate between individual events and thus decrease the applied field. This leads to the typical local gain drop of MCPs at high count rates and operation near saturation, in which the positive space charges in the walls compensate the applied electrical field.

\begin{figure}
	\centering
		\includegraphics[width=.7\textwidth]{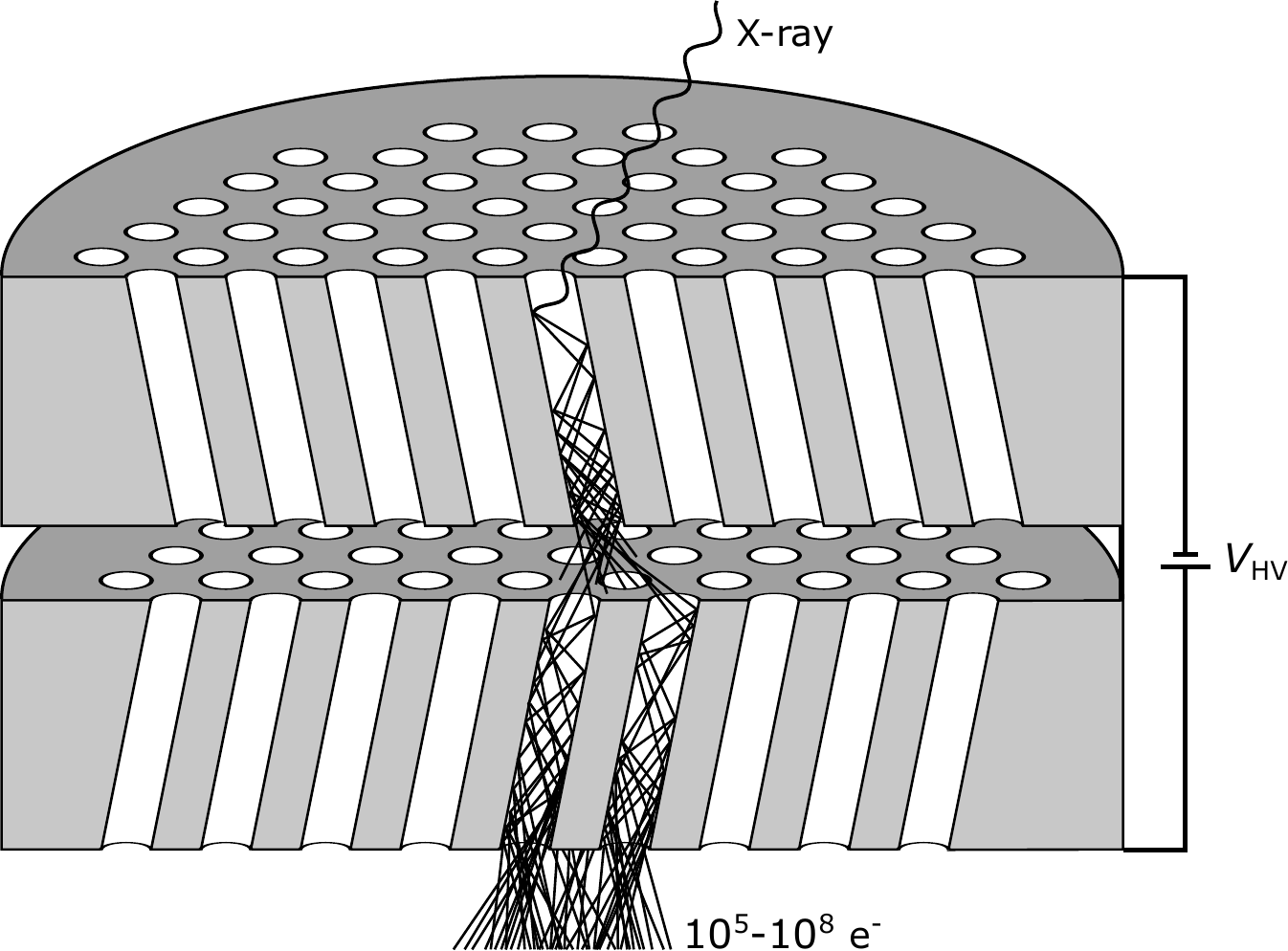}
	\caption{\label{fig:mcp_stack}
  Cross-section of a stack of two MCPs. Photoelectrons, which are produced when X-rays interact with the top plate, are accelerated inside the channels by the applied high voltage. When impacting the channel wall an avalanche of secondary electrons is generated, finally forming a charge cloud of typically $10^5$ -- $10^8$ electrons.}
\end{figure}

Most MCPs have straight channels which are inclined with respect to the surface normal around 10\degree. This inclination increases the response to photons incident normal to the surface, but more important it allows to suppress ion feedback, particularly when more than one plate is operated in a detector. Typical detector configurations feature either two plates -- the so-called \textit{Chevron stack} -- or three plates -- the \textit{Z-stack}. The direction of the channel inclination is inverted in each plate, as shown in Figure~\ref{fig:mcp_stack} for a stack of two MCPs. While a single plate is limited to a gain of about $10^4$ before reaching saturation, a stack of two plates achieves up to $10^8$ and a three plates stack can even exceed $10^9$.

\begin{figure}
	\centering
		\includegraphics[width=.6\textwidth]{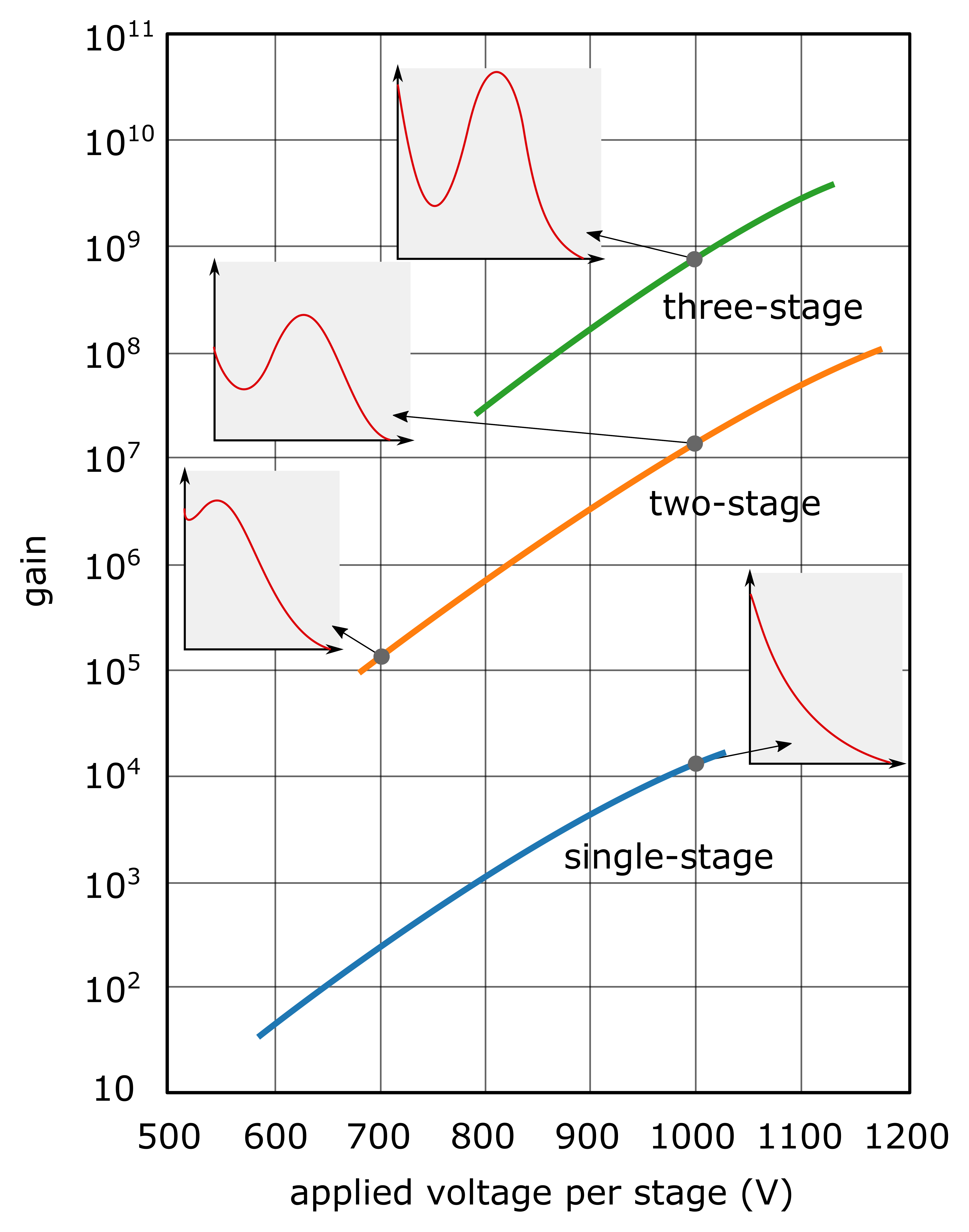}
	\caption{\label{fig:mcp_charge}
  Generic gain dependence on the applied voltage for a single-stage, a stack of two \textit{(Chevron stack)}, and a stack of three MCPs \textit{(Z-stack)}. The inlayed plots illustrate the typical shape of the actual charge distribution at the points on the gain curves (probability vs. number of electrons in an event).}
\end{figure}

The necessary gain strongly depends on the application -- particularly for imaging purposes on the type of position-sensitive anode and the required position resolution. The generic gain-over-voltage characteristics of the three common plate configurations are shown in Figure~\ref{fig:mcp_charge}. A key quantity to exploit the photon-counting capabilities of an MCP detector is the charge distribution. For a photon-counting operation without any readout noise the event threshold has to be set above the electronics noise level of the system without cutting away photon events and thus sacrificing quantum efficiency. This is achieved best when a clear separation between the electronic noise charge (ENC) of the readout system and the charge distribution of real events is achieved. The inlay plots in Figure~\ref{fig:mcp_charge}display the typical charge distributions under different operation conditions and show that stack of at least two MCPs is needed to fulfill this requirement, since the charge distribution of a single MCP is an exponential without a peak structure for photon induced events.

A special case are MCPs with curved or J-like channels that also reduce ion feedback and allow photon-counting operation in single-stage mode \citep{Timothy1977}. Such plates were mainly applied in \textit{MAMA (Multi Anode Microchannel Array)} detectors that achieve with a rather simple digital-like readout a high spatial resolution. Such detectors have been applied in the NUV and FUV channel of STIS (Space Telescope Imaging Spectrograph) as well as in other instruments of the Hubble space telescope (HST) \citep{Woodgate1998,Timothy2016}.

Similarly to CEMs, MCPs can only be operated under high vacuum conditions for several reasons:
\begin{enumerate}
	\item Under atmospheric conditions the mean free path for electrons would not be sufficient to reach and impact the opposite channel wall and thus trigger secondary electron emission.
	\item If the pressure is not sufficiently low a large number of positive ions are created within the channels and are accelerated towards the front face of the plate. Due to their higher mass they move much slower than electrons and when they impact the channel wall, they trigger the formation of additional delayed pulses -- the so-called \textit{ion feedback}.
	\item Most photocathode materials that are coated on the top face to increase the sensitivity degrade rapidly under atmospheric conditions. Some are hygroscopic, others degrade under exposure to oxygen. Depending on the specific material the time constant for this process varies from seconds to hours.
	\item Applying typically $1\,\mathrm{kV}$ to a millimeter thick plate is only possible under high vacuum without risking corona discharges that would immediately destroy the MCPs. Since the breakthrough voltage of air and other gases intially drops when lowering the pressure and reaches a minimum in the $1\,\mathrm{mbar}$ regime, a pressure below $10^{-3}\,\text{mbar}$ is necessary for a safe and stable operation of MCPs.
\end{enumerate}



\subsection{\textit{Quantum detection efficiency}}

The \textit{quantum detection efficiency (QDE)} is one of the basic benchmark parameters for a radiation detector. It is defined as the probability that a single photon hitting the sensitive detector area is registered and processed by the electronics. Care has to be taken because literature quite often refers only to quantum efficiency (QE) without clarifying what is meant exactly. In particular, this can be misleading when comparing photon-integrating detectors to photon-counting detectors like MCPs.

One of the primary parameters defining the QDE of an MCP detector is the \textit{open area ratio (OAR)} -- the ratio of open channel surface to the total surface of the plate. For conventional lead glass MCPs, the lead content in the bulk material of up to 48\,\% by weight is mostly responsible for the hard X-ray and gamma-ray response, while soft X-rays and UV photons interact only with the roughly 10\,nm thick semiconducting surface layer that is depleted from lead and enriched with lighter elements \citep{Fraser2009}. Therefore, particular for the softer part of the X-ray spectrum, an increased OAR should lead to a higher QDE. With small straight channels of a few micrometer in diameter an OAR of up to roughly 70\,\% can be reached. By introducing in the fabrication process a second etching step that affects mainly the front-side channel openings, a funnel shape can be created that can increase the OAR to more than 90\,\%.

Another more common method to increase the QDE is to place an \textit{electron repeller grid} -- usually a thin wire mesh or metal foil -- in front of the detector to which a negative bias voltage is applied with respect to the front surface of the first MCP (cf. Figure~\ref{fig:chandra}). When a photon hits the bulk material between the channels and the associated photoelectron is emitted away from the detector, the electric field bents the electron back towards the MCP where it may again hit the bulk and be lost or enter a channel and be detected. Thickness and distance of the grid wires as well as the applied voltage are a trade-off between optimizing the field and thus the position resolution and reducing shadowing effects.

Uncoated MCPs have in the EUV and soft X-ray bands a QDE in the range of 1-10\,\%, typically decreasing with energy \citep{Fraser2009}. Additionally, a strong dependance on the angle of incidence is seen, with lower efficiencies at close to normal as well as grazing incidence and peaking close to the critical angle of reflection from the MCP glass. This empirical finding is in accordance with theoretical predictions \citep{Bjorkholm1977,Fraser1982}. The QDE can be enhanced by coating a photocathode on the front surface that extends into the beginning of the channels. While the photocathode material of choice for the first soft X-ray detectors was magnesium fluoride (MgF\textsubscript{2}) that was mainly chosen for its stability, cesium iodide (CsI) became later the baseline material and was commonly used in a large waveband from 180\,nm down to 0.2\,nm. A drawback is the hygroscopic nature of CsI leading to degradation within minutes when handled under atmospheric conditions. Even water molecules gettered in the bulk of an MCP can induce this degradation process, therefore the plate should be baked out thoroughly before coating \citep{Fraser1982}. A more stable option that reaches even higher efficiencies between 150\,nm and 4\,nm is potassium bromide (KBr). An MCP with a KBr photocathode can be handled in air for about half an hour without considerable negative effects and even longer when flushed with dry nitrogen. Cesium bromide (CsBr) is another option for the short wavelengths between 10\,nm and 2\,nm.

Besides an increased QDE in the detection waveband a photocathode offers the possibility to avoid background from longer wavelength by choosing a material with a sufficiently wide band gap. This allows the fabrication of solar-blind detectors with a vanishing sensitivity for visible light. For EUV and X-ray detectors the Lyman-Alpha line at 121.6\,nm is often a considerable source of background and even typical wide-bandgap photocathodes have a finite sensitivity at this wavelength, thus a thin metal filter is usually foreseen to remove this out of band background together with low energy charged particles.

Particularly in the NUV and down to about 110\,nm in the FUV band, \textit{semitransparent photocathodes} coated on the inside of a window are an interesting alternative to thick, opaque photocathodes directly on the MCP. With a stable and thick window -- MgF\textsubscript{2} for the FUV, quartz for the NUV -- a sealed detector can be build that does not need any shutter mechanism to keep MCPs and photocathode under vacuum or in a protective gas atmosphere before it can be opened in space \citep{Siegmund2021,Conti2022}. This is not possible for EUV and X-ray detectors as no material with sufficiently high transmission for these wavelengths exist that is leak tight and stable to cope with the pressure difference. However, semitransparent photocathodes on X-ray transparent layers were investigated and used for some time in combination with shutter mechanisms, but it could be proven that these are not more efficient than conventional opaque photocathodes while being less reproducible \citep{Fraser2009}. In contrast to opaque photocathodes, the thickness is a critical parameter in this \textit{semitransparent mode}, since the layer must be thick enough to absorb the photon but at the same time thin enough to allow the emission of the photoelectron.

While channel pitch $p$ and diameter $d$ are not relevant for the sensitivity for soft X-rays, the detection shifts to the bulk and the photoelectron paths are longer when the photon energy increases, and thus a thick plate with thin walls, i.e. large length $l$ and small $p-d$, is optimal for hard X-rays. However, while around 20\,keV up to 10\,\% peak QDE can be achieved, the efficiency stays at higher energies up to 1\,MeV quite constant below 4\,\% \citep{Fraser2009}. Therefore, technologies with tremendously higher QDE than MCPs exist for this range.

\subsection{\textit{Position-sensitive readout, spatial and temporal resolution}}

Historically, MCP detectors were the first detectors for soft X-ray astronomy featuring a high spatial resolution significantly below $100\,\mathrm{\mu m}$. The ultimate limit for single events is the channel pitch, and with modern readouts the question is not how to reach this but instead how low the power consumption can be -- to facilitate the thermal design of the instrument -- and how low the gain can be chosen -- to enlarge the MCP lifetime. However, for a statistical sample of $n$ counts in a particular feature, a continuously digitally oversampled detector can reach a spatial FWHM that is even better than the channel pitch by a factor $\sqrt n$ \citep{Siegmund1986}.

\begin{figure}[t!]
	\centering
	\subfigure[Wedge-and-strip anode (WSA)]{
  	\includegraphics[width=.51\textwidth]{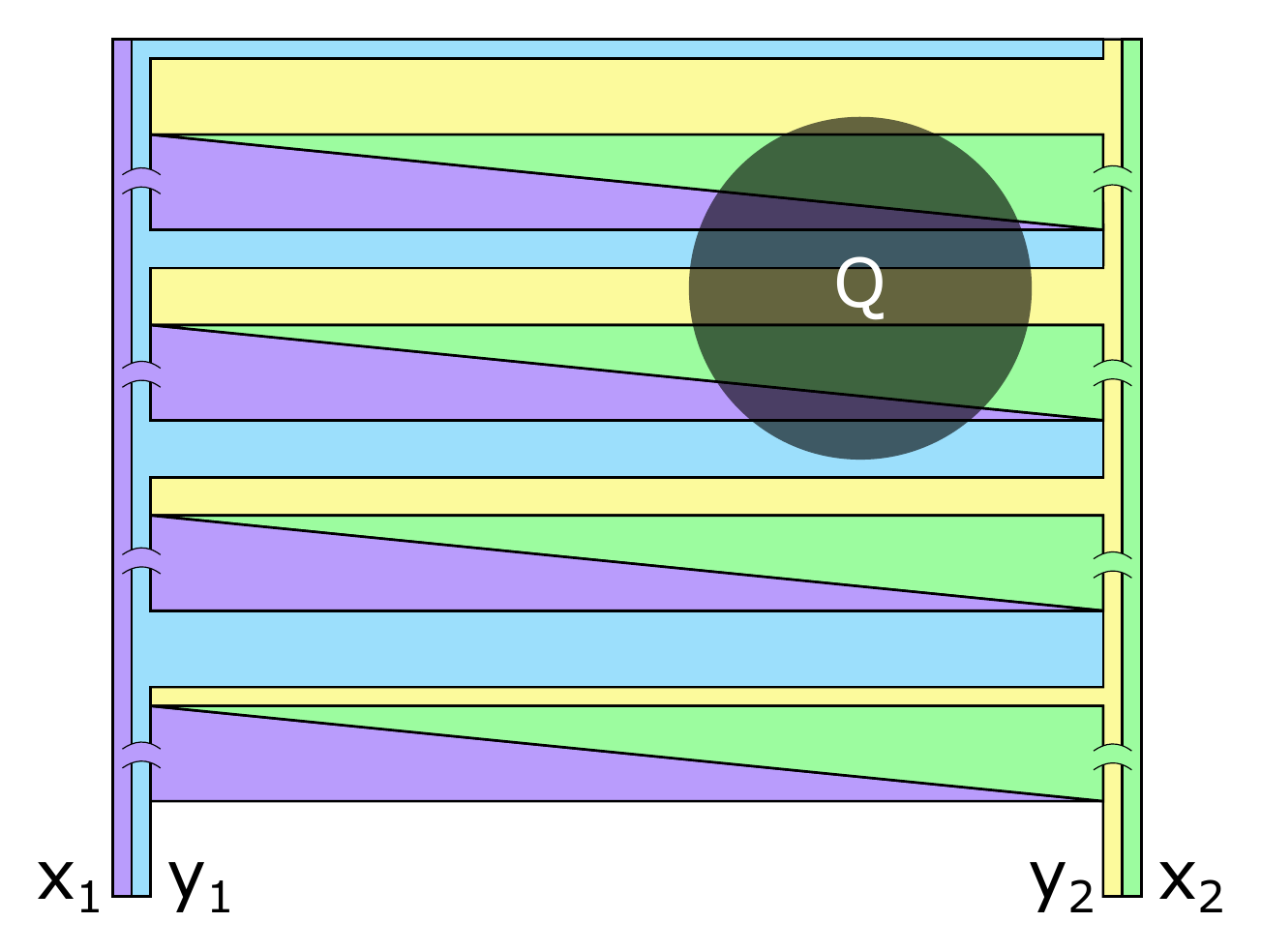}
		\label{fig:WSA}
	}
  \subfigure[Crossed delay-line anode (DLA)]{
  	\includegraphics[width=.45\textwidth]{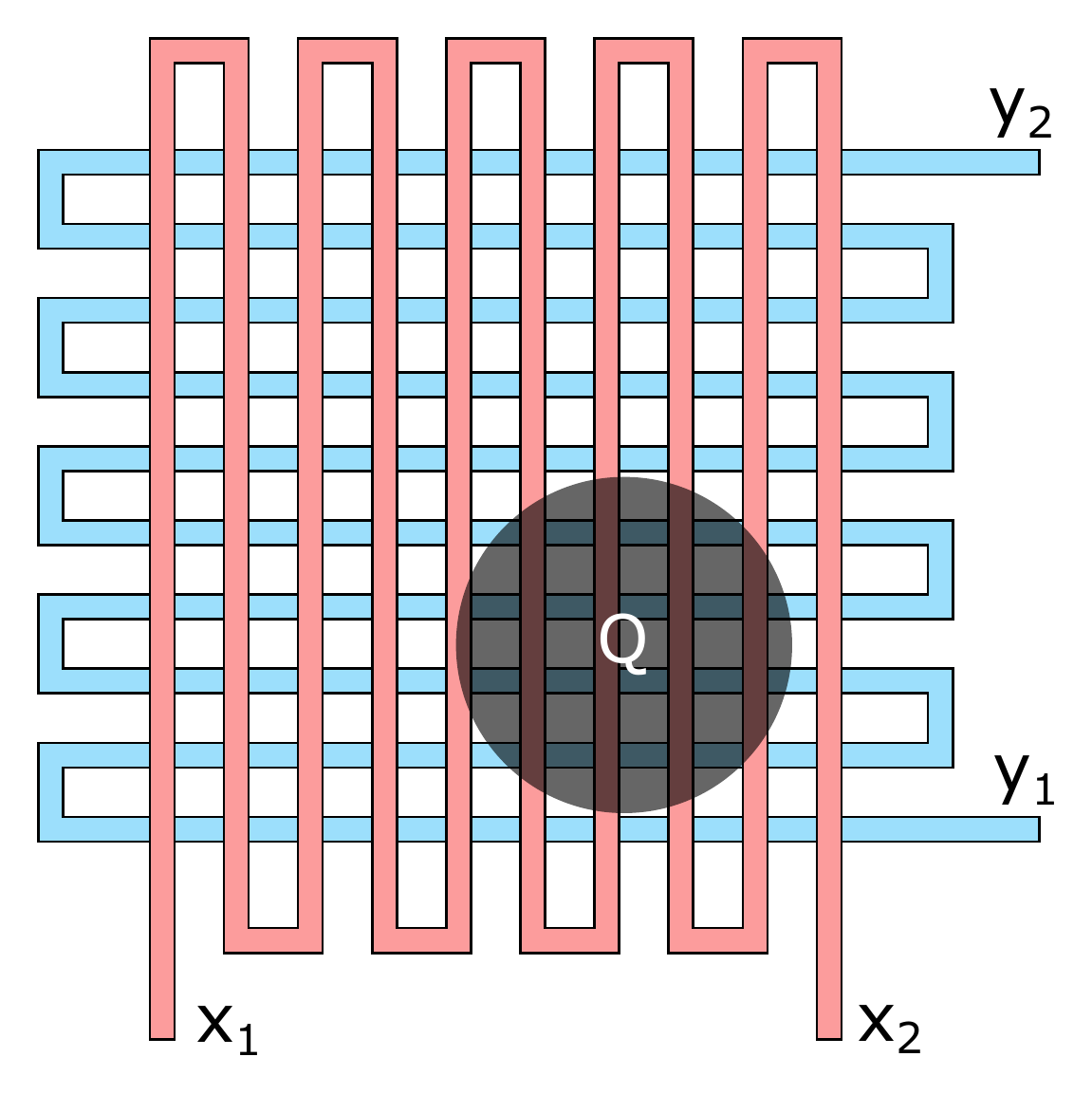}
		\label{fig:DLA}
	}
	\subfigure[Co-planar cross-strip anode (CSA)]{
  	\includegraphics[width=.54\textwidth]{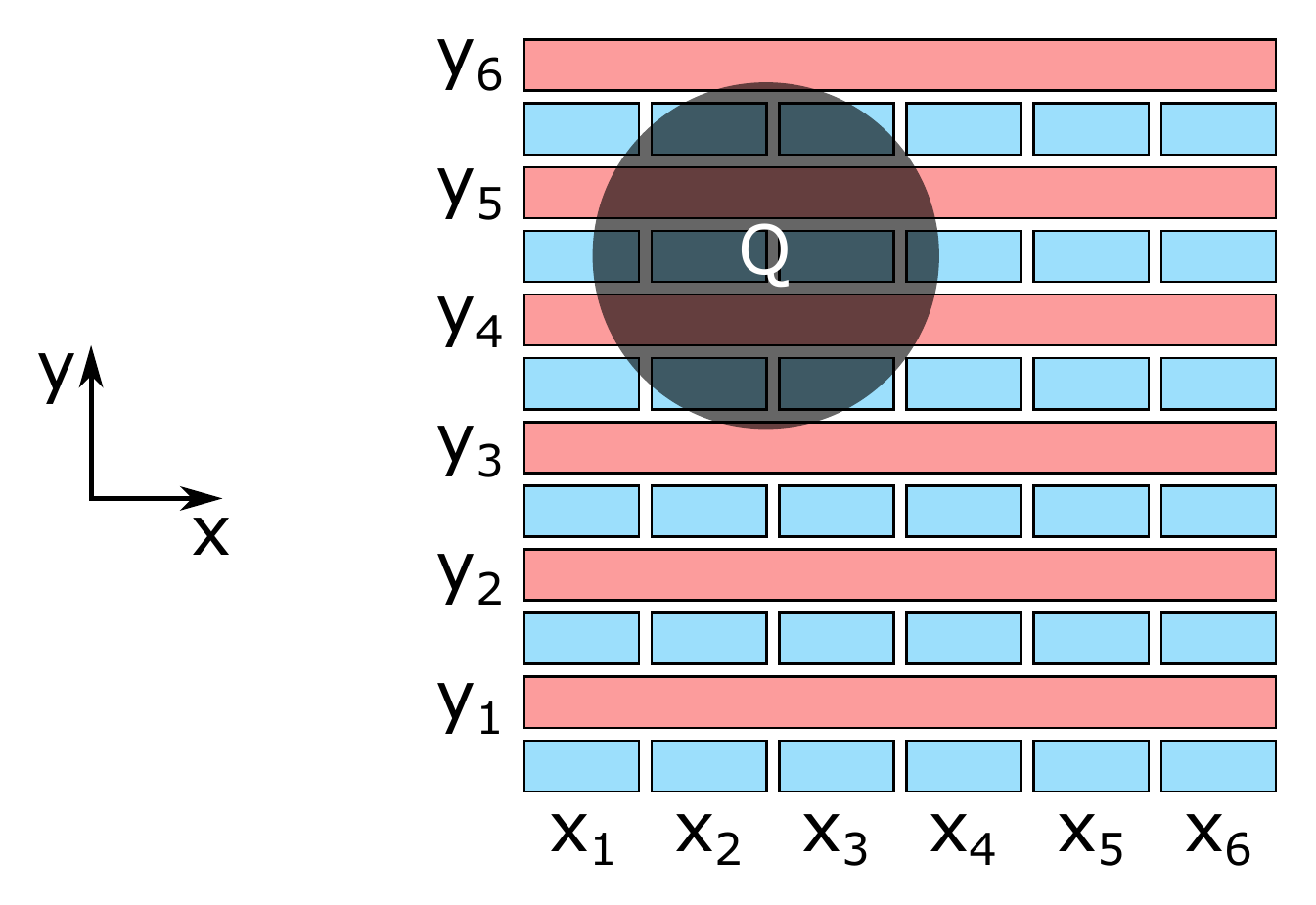}
		\label{fig:CSA}
	}
	\caption{\label{fig:anodes}
  Schematic drawings of typical position-sensitive anodes for MCP detectors: (a) Wedge-and-strip anode: the position is encoded in the charge shared between two complementary wedges for the x-direction and two strips with complementary width in the y-direction. (b) Crossed delay-line anode: the difference in time of arrival on the different ends of the two delay-lines is used to determine the position information. (c) Co-planar cross-strip anode: the charge is recorded with a large number of parallel strips in x- and y-direction with individual readout channels. The position is determined with a resolution considerably smaller than the strip pitch by applying a centroiding algorithm to the digitized data.}
\end{figure}

In principle, the same position-sensitive anode designs can be implemented in MCP detectors that are also used for imaging proportional counters (s. Section~\ref{sec:imaging_counters}). A revolutionary anode for MCP detectors was the \textit{wedge-and-strip anode (WSA)} \citep{Martin1981} that allowed with the principle of charge sharing spatial resolutions down to about $40\,\mathrm{\mu m}$ with only four readout channels. The WSA geometry is sketched in Figure~\ref{fig:WSA}. The x- and y-coordinates of the center-of-gravity of the charge cloud can be retrieved with rather simple calculations:

\begin{equation}
x=\frac{x_2}{x_1+x_2}
\label{eq:WSAx}
\end{equation}

\begin{equation}
y=\frac{y_2}{y_1+y_2}
\label{eq:WSAy}
\end{equation}

In an improved design in which two of the four electrodes are combined the number of channels even reduces to three by the price of increasing the capacitance per channel. In general, the signal-to-noise ratio of a charge-sensitive preamplifier increases with decreasing input capacitance and, therefore, the number of readout channels is usually a trade-off between noise on the individual channels and the number of channels.

However, similar to the position-sensitive readout of proportional counters, also for MCP detectors an alternative concept to charge-sharing uses the time-of-arrival of signals rather than the charge-spreading information to determine the position. In later astronomical MCP detectors this principle of the \textit{delay-line anode (DLA)} clearly dominated. It is sketched in Figure~\ref{fig:DLA}. Two delay-lines are routed meander-like over the anode area and \textit{time-to-digital converters (TDC)} on both ends of each line sense the time-of-arrival of a charge pulse. From the time differences the x- and y- positions can be calculated. Depending on the geometry, the MCP gain, and the complexity of the readout electronics a DLA can achieve a position resolution below $10\,\mathrm{\mu m}$ and thus allows imaging of the MCP pores.

Two other anode concepts under development for MCP detectors promise some advantages over the established ones but have not yet been applied in an astronomical instrument: the co-planar \textit{cross-strip anode} (CSA, s. Figure~\ref{fig:CSA}, \cite{Conti2018}) and the direct readout of the charge cloud with an ASIC \citep{Tremsin2020}. Both point in the same direction as was already sketched above for micropattern readout anodes for proportional counters (s. Section~\ref{ssec:micropattern-gas}) and can only be realized with modern high density integrated circuits that feature a large number of channels and sufficient radiation hardness. The CSA employs a large number of linear electrodes connected to individual readout channels and by applying a centroid algorithm on the digitized charge information a position resolution considerably smaller than the strip-pitch can be reached \citep{Conti2018}.

Besides a high spatial resolution, the MCP technology offers the possibility for astonishing high temporal resolution compared to gas or semiconducting X-ray detectors, but this feature was never exploited in astronomical detectors. The signal transit time through an MCP is typically around $100\,\mathrm{ps}$ with a jitter at least a factor of 10 lower and scaling with the channel diameter $d$. Therefore, MCP detectors offer an ultimate temporal resolution of the order of $10\,\mathrm{ps}$. A rather simple method to reach a high temporal resolution without sacrificing spatial resolution is to decouple these quantities: a fast timing of the order of $500\,\mathrm{ps}$ can be reached by triggering from the pulse on the MCP backside electrode that is generated when the charge cloud exits, while the event position is encoded with one of the slow readouts already discussed.

Usually an MCP detector does not feature any energy resolution besides its out-of-band rejection. However, a poor but finite energy resolution is possible for soft X-rays if the first plate of a stack is operated with lower than saturation voltage. In this configuration slight gain variations depending on the X-ray energy are seen. This would make a two to three color photometry possible, e.g. with the HRC on Chandra (s. Section~\ref{ssec:MCP_appl}), but has not been exploited in any X-ray mission.

\subsection{\textit{Applications in EUV and X-ray astronomy}}
\label{ssec:MCP_appl}

MCP detectors were for many years the preferred detector technology for UV instrumentation, particularly in the FUV below $180\,\mathrm{nm}$. They were used on many missions, e.g. IUE (International Ultraviolet Explorer; launched in 1978), HST (Hubble Space Telescope; launched in 1991, four service missions), ORFEUS (Orbiting and Retrievable Far and EUV Satellite; flown in 1993 and 1996 with NASA space shuttles), FUSE (Far Ultraviolet Spectroscopic Explorer; launched in 1999), and GALEX (Galaxy Evolution Explorer, launched in 2003), to name just the most important ones for stellar astronomy.

In the EUV wavelength band from the Lyman limit at $91.2\,\mathrm{nm}$ down to about $10\,\mathrm{nm}$, hardly an alternative to MCPs exist. A recommendable overview over instrumentation and science topics in this waveband is given in \citet{Barstow2003}, later complemented to account for more recent developments in \citet{Barstow2014}. Two major surveys were carried out in the EUV, both with instruments based on MCP detectors: the first one with the EUV camera (WFI -- Wide Field Imager) on ROSAT (ROentgen SATellite; launched in 1990), and the second one with EUVE (Extreme Ultraviolet Explorer; launched in 1992), the first mission fully dedicated to the EUV waveband \citep{Bowyer1991}. And even today, particularly when a photon-counting detector is needed, the baseline technology in the EUV remains MCPs with a KBr photocathode, as planned for the proposed ESCAPE \citep{France2022} and SIRIUS missions \citep{Barstow2012}. 

There have been several missions in the soft X-ray band that employed MCP-based detectors, often combined with other instruments and detector technologies on the same satellite. The following compilation is surely not complete, but tries to give an overview of missions and instruments with an outstanding impact on the field.

A cornerstone mission for X-ray astronomy was the Einstein (HEAO-2) observatory launched in 1978 \citep{Giacconi1979}. It was not only the first imaging X-ray telescope in space, but also the first mission to apply an MCP detector in its High Resolution Imager (HRI) \citep{Kellogg1976}. Observations with this instrument remained for many years the highest resolution images of the X-ray sky. A quite recommendable recount of the development history of this mission was published in \citet{Tucker2013}. The HRI instrument has been tested on several rocket flights and finally after four failures it worked flawlessly in the fifth test flight only four months prior to launch. In space it worked without problems and produced images in the 200\,eV to 4\,keV band. The detector was enclosed in a vacuum housing that was pumped with an ion pump before a door was opened in space to expose the MgF\textsubscript{2} coated sensitive front. In fact, HEAO-2 featured three HRIs, identical but for the composition of their UV/ion shields.

In 1990 the German-led ROSAT mission was launched with one of the instruments onboard going by the same name HRI. This is not a coincidence since this US contribution to the mission was almost identical to the HRI design for HEAO-2, only the MgF\textsubscript{2} photocathode was exchanged for CsI. While the spatial resolution stayed the same, this led to a boost in efficiency by a factor 1.5 to 4 over the spectral range of the instrument of 0.1--2 keV \citep{Pfeffermann1987}.

With its Wide Field Camera (WFI) contributed by the UK ROSAT featured a second instrument read out with an MCP detector. It was sensitive in the EUV range 6--20\,nm with a field-of-view (FOV) \ang{30;;} in diameter. The MCP detector for the WFI applied also a CsI photocathode similar to the HRI detector, but with an additional electron repeller grid in front. A further sensitivity enhancement of up to a factor of 2.5 at the edges of the FOV was reached by matching the spherical shape of the focal surface with curved MCPs \citep{Barstow1985}.

ROSAT was designed for a minimal lifetime of 18 months with a goal of five years, but actually the mission ran for almost nine years. An incident a few months before the mission clearly demonstrated the severe damage that an overexposure can create in an astronomical sensor: due to a reaction wheel failure the HRI was exposed to sunlight and afterwards hardly usable \citep{Truemper1999}. However, disregarding this particular event, it is notable that the MCP-based detectors were still fully operational when the proportional counters of ROSAT already became inoperable because of gas loss and thus insufficient pressure.

\begin{figure}
	\centering
		\includegraphics[width=\textwidth]{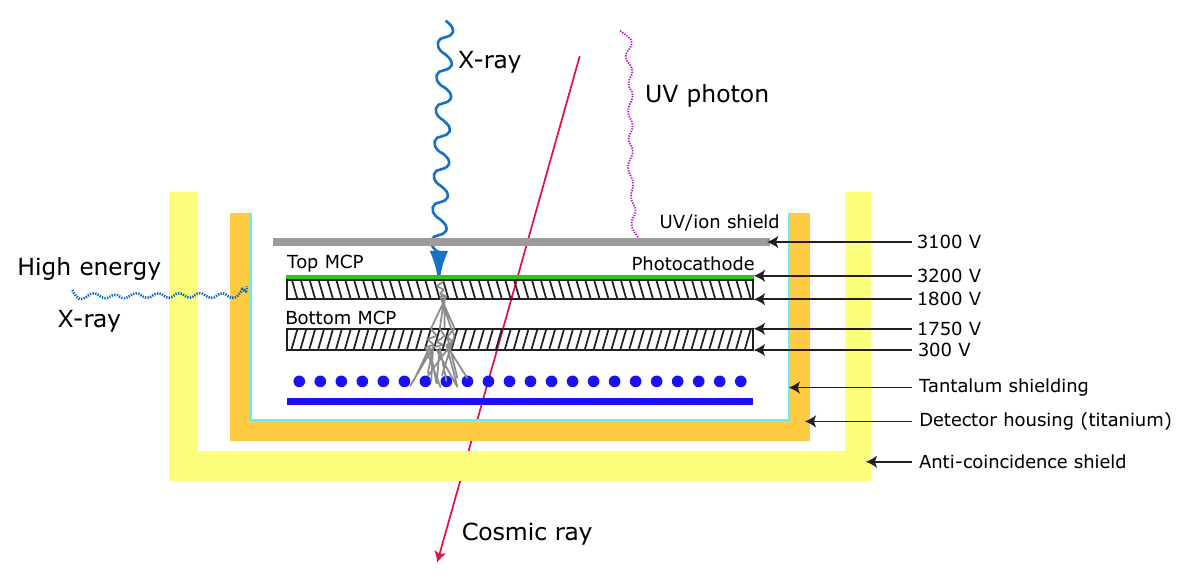}
	\caption{\label{fig:chandra}
  Cross-sectional sketch of the Chandra HRC-I detector. A stack of two square $10 \times 10\,\mathrm{cm^2}$ MCPs is used with a thick CsI photocathode coated on the top plate. The exiting charge is detected with a crossed wire grid that allows a position resolution well matched to the angular resolution of the X-ray optics. A thin metallic filter provides shielding from UV photons and ions. Further, to enhance the sensitivity, this filter is put on a negative potential with respect to the photocathode to deflect photoelectrons emitted from the surface back to the MCPs. The inner part of the detector is shielded on five sides: a passive shielding catches high energy X-rays and an active anti-coincidence system vetos events induced by high energy particles.}
\end{figure}

Also the High Resolution Camera (HRC) of the Chandra X-ray Observatory is a direct descendant of the Einstein HRI. The major improvements are the twenty times larger image area while maintaining the same spatial resolution, a thick CsI photocathode with further enhanced sensitivity, and a partially saturated operation of the MCP stack that allowed in principle a finite energy resolution enabling two to three color photometry \citep{Weisskopf2003}. A sketch of the HRC-I imaging detector including its housing is shown in Figure~\ref{fig:chandra}. With a sensitive area of $10 \times 10\,\mathrm{cm^2}$ it is the largest MCP detector ever used in X-ray astronomy. In order to reduce the background level of this particularly large sensitive volume the inner part of the detector is shielded with the high-Z material tantalum, primarily to capture penetrating high energy photons. Additionally, the sensitive volume is surrounded by an active anti-coincidence shield so that penetrating highly energetic charged particles are vetoed.

\section{\textit{Future prospects}}
\label{sec:prospects}

Most current developments in the MCP domain focus on applications in the FUV and EUV bands where MCPs still remains the dominating technology for a wide range of astronomical applications \citep{France2022}. In the soft X-ray band above 200\,eV the efforts concentrate on semiconductors, e.g. silicon drift detectors (SDDs) for large sensitive areas \citep{Rachevski2014,Evangelista2018} or active pixel sensors like depleted field effect transistor arrays (DepFETs) \citep{MuellerSeidlitz2022}, as well as on micro-calorimeter arrays with their unprecedented energy resolution \citep{Pajot2018,Taralli2022}. On the other wavelength end, silicon detectors like CCDs or CMOS sensors are improved towards lower wavelengths by anti-reflective coatings and delta-doping so that they are superior to MCPs in the NUV for a large number of applications.

Concerning local count rate capabilities the MCP technology has reached its principle limit since many years. The recharging time constant depends on the plate resistance that cannot be decreased without risking a thermal run-away. Concerning sensitivity, there is no prospect of substantial improvements in the EUV and soft X-ray band, while for NUV and FUV new photocathode materials like GaN and AlGaN are actively developed \citep{Conti2020,Cai2021}. An enormous progress was made in the last years concerning lifetime and gain stability with the development of borosilicate MCPs that are fabricated with the hollow-tube technique and functionalized with an ALD process \citep{Cremer2020}.

Substantial efforts are also invested in compact and low-power readout electronics for cross-strip anodes that allow high spatial resolution with lower MCP gain as for the baseline cross delay-line anodes \citep{Pfeifer2014,Siegmund2020}. Another readout that is still under active development is the direct readout with an ASIC. Most advanced are here the efforts to use a Medipix/Timepix chip that was originally developed for high-energy detector applications \citep{Tremsin2020}.

An interesting secondary application of the MCP technology is in so-called lobster eye X-ray optics, a wide field-of-view X-ray imaging technique invented already in the 1990s \citep{Fraser1992} but realized only now in several upcoming missions: SMILE, SVOM, and Einstein Probe. All these missions are presented in the section \textit{X-ray Missions} within this handbook.

In the field of gaseous detectors for X-ray astronomy, during the last years the development focused on gas pixel detectors for polarimetric instruments. Although several attempts with instruments on rockets and satellites have been made over the last 50 years in this domain, only the recent advancements with gas pixel detector technology allow to determine polarization degree and direction of astrophysical sources over a wide energy range and with a high signal-to-noise ratio. While IXPE is already applying this concept in space, the upcoming eXTP mission is designed to be able to observe an even larger number of sources for which the polarimetric information will be of paramount importance to better constrain the geometry and the underlying physics.

\section*{\textit{Cross-References}}

Several chapters in this handbook are dedicated to missions applying proportional counters:
\begin{itemize}
	\item \textit{The AstroSat Observatory} by Singh, K.
  \item \textit{eXTP} by Santangelo, A., Nan Zhang, S.
  \item \textit{MAXI: Monitor of All-sky X-ray Image} by Mihara, T., Tsunemi, H., Negoro, H.
  \item \textit{IXPE: The Imaging X-ray Polarimetry Explorer} by Weisskopf, M., Soffitta, P., Ramsey, B., Baldini, L.
\end{itemize}

More information on advanced gaseous detectors specifically for measuring X-ray polarization can be found in these chapters:
\begin{itemize}
	\item \textit{Gas pixel detectors for polarimetry} by Soffitta, P., Costa, E.
	\item \textit{Time projection chamber X-ray Polarimeters} by Black, K., Zajczyk, A.
\end{itemize}

Microchannel plate detectors are used on the missions in the follow chapters:
\begin{itemize}
	\item \textit{The AstroSat Observatory} by Singh, K.
  \item \textit{The Chandra X-ray Observatory} by Wilkes, B., Tananbaum, H.
\end{itemize}

Microchannel plates as collimators for so-called \textit{lobster eye} wide field X-ray telescopes are applied in the missions in the follow chapters:
\begin{itemize}
	\item \textit{The Einstein Probe Mission} by Yuan, W., Zhang, C., Chen, Y., Ling, Z.
  \item \textit{The SMILE mission} by Branduardi-Raymont, G., Wang, C.
	\item \textit{SVOM} by Wei, J.
\end{itemize}


\printbibliography

\end{document}